\documentclass[12pt]{article}

\usepackage{epsfig,graphics,amssymb,here}

\newcommand{\be} {\begin{equation}}
\newcommand{\ee} {\end{equation}}
\newcommand{\bma} {\begin{math}}
\newcommand{\ema} {\end{math}}
\newcommand{\beqa} {\begin{eqnarray}}
\newcommand{\eeqa} {\end{eqnarray}}

\def\vec#1{\mathchoice{\mbox{\boldmath$\mathrm\displaystyle#1$}}
{\mbox{\boldmath$\mathrm\textstyle#1$}}
{\mbox{\boldmath$\mathrm\scriptstyle#1$}}
{\mbox{\boldmath$\mathrm\scriptscriptstyle#1$}}}
\newcommand{\bm}[1]{\mbox{\boldmath$#1$}}  %fette Mathe-Symbole
\renewcommand{\vec}{\bm}
\newcommand{\simgt}{\hbox{ \raise3pt\hbox to 0pt{$>$}
    \raise-3pt\hbox{$\sim$} }}
\newcommand{\simsm}{\hbox{ \raise3pt\hbox to 0pt{$<$}
    \raise-3pt\hbox{$\sim$} }}
\begin{document}

\begin{flushright}
HD--THEP-99-53 \\
%\end{flushright}
%\begin{flushright}
M/C--TH-99/19 \\
%\end{flushright}
%\begin{flushright}
LPT-ORSAY-99-106 
\end{flushright}

\vspace{0.2cm}

\begin{center}

{\LARGE{Observing the Odderon: Tensor Meson 
Photoproduction}}\footnote{
\hphantom{dfde}$email\; adresses$: 
  berger@lyre.th.u-psud.fr,
  ad@a13.ph.man.ac.uk,\\
  \hphantom{dfde}h.g.dosch@thphys.uni-heidelberg.de,
  o.nachtmann@thphys.uni-heidelberg.de}

\vspace{0.1cm}

\end{center}

\begin{center}
{\large 
E. R. Berger${}^{\rm a}$, 
A. Donnachie${}^{\rm b}$,
H. G. Dosch${}^{\rm c}$,
O. Nachtmann${}^{\rm c}$ }
\end{center}

\vspace{0.1cm}

$\hphantom{ss\;ss}{}^{{\rm a}}$ : 
$\, \;$ LAPTH, Universite Paris-Sud, Batiment 211,\\ 
        \hphantom{ : ${}^{1,3,5}$ssss}
        F-91405 Orsay Cedex, France

$\hphantom{ss\;ss}{}^{{\rm b}}$ : 
$\, \;$ Department of Physics and Astronomy,\\ 
        \hphantom{ : ${}^{1,3,5}$ssss}
        University of Manchester, Manchester M13 9PL, UK

$\hphantom{ss\;ss}{}^{{\rm c}}$ : 
$\, \;$ Institut  f\"ur  Theoretische Physik
        der Universit\"at Heidelberg,\\ 
        \hphantom{ : ${}^{1,3,5}$ssss}
        Philosophenweg 16, 
        D-69120 Heidelberg, Germany

\vspace{0.5cm}

\thispagestyle{empty}

\begin{abstract}

We calculate high-energy photoproduction of the tensor meson 
$f_2(1270)$ by odderon and photon exchange in the reaction 
$\gamma + {\rm{p}} \rightarrow f_2(1270) + {\rm{X}}$, where X is 
either the nucleon or the sum of the N(1520) and N(1535) baryon
resonances. Odderon exchange dominates except at very small transverse
momentum, and we find a cross section of about 20 nb at a centre-of-mass
energy of 20 GeV. 
This result is compared with what is currently known experimentally
about $f_2$ photoproduction. We conclude that odderon exchange is not
ruled out by present data. On the contrary, an odderon-induced cross 
section of the above magnitude may help to explain a puzzling result 
observed by the E687 experiment.

\end{abstract}

\newpage

\section{Introduction}

In this paper we investigate the high-energy diffractive production of the 
tensor meson $f_2(1270)$ by real and virtual photons. Specifically we study
$\gamma^{(*)}{\rm p} \rightarrow f_2 {\rm X}$, where X is either the
proton or
the sum of 
the negative parity N(1520) and N(1535) resonances. For the latter two it is 
expected that the dominant exchange is the nonperturbative odderon 
\cite{nico1}, the $C = P = -1$ partner of the pomeron \cite{etal}. 
It has been shown 
\cite{doruod}
that the suppression of the odderon contribution in pp ($\bar{\rm p}$p)
scattering can be explained by diquark clustering in the proton, provided
the diquark is sufficiently small $\le 0.3$ fm. This
suppression does not operate if the nucleon dissociates into a negative
parity state \cite{donaru}. 
This paper is a continuation of the work presented in 
\cite{donaru,etal} where the electroproduction of pseudoscalar mesons, in 
particular the reaction $\gamma^{(*)}{\rm p} \rightarrow \pi^0 {\rm X}$ was 
studied. An explanation of the general philosophy of our approach and 
references to related work can be found there. A recent review about 
odderon physics is \cite{nico2}.

\begin{figure}[htb]
  \unitlength1.0cm
  \begin{center}
    \begin{picture}(15.,8.8)

      \put(-0.6,1.0){
        \epsfysize=4.0cm
        \epsffile{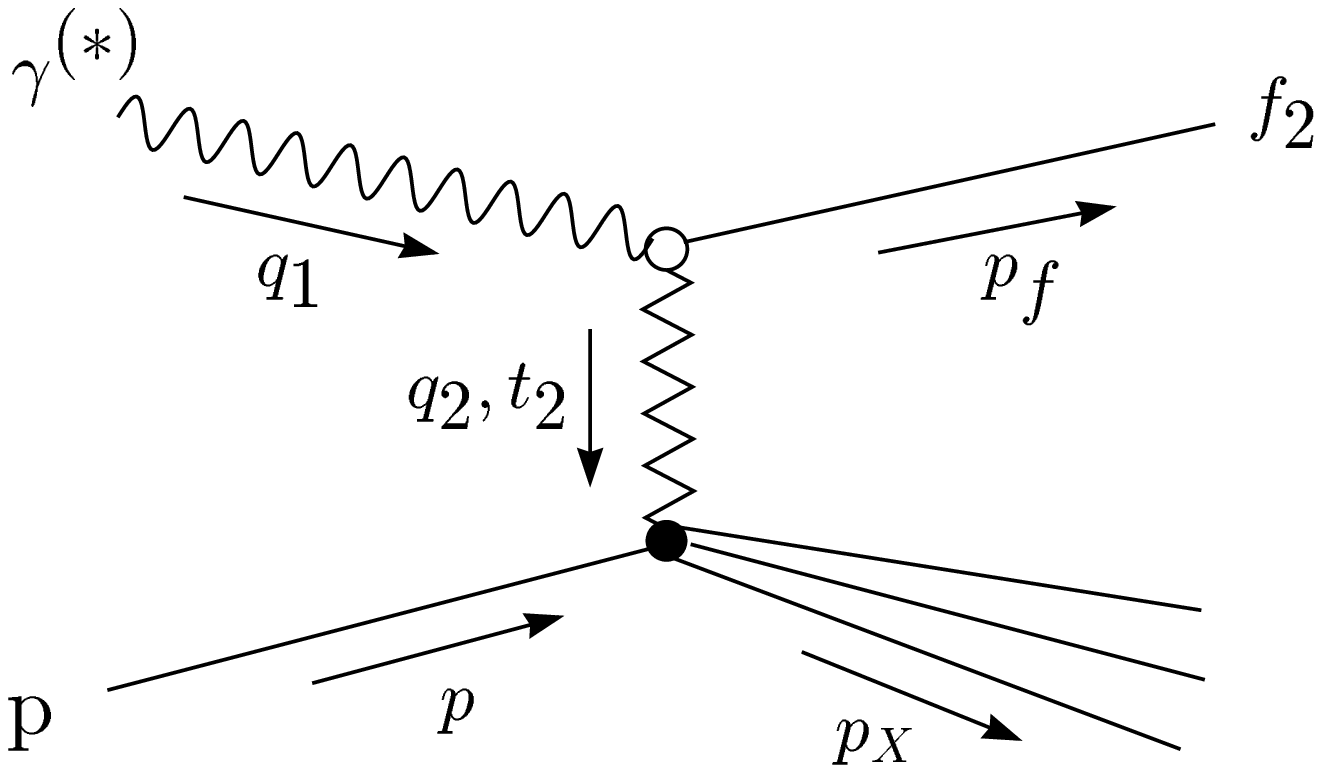}}

      \put(0.4,0){(a)}

      \put(7.7,1.0){
        \epsfysize=4.0cm
        \epsffile{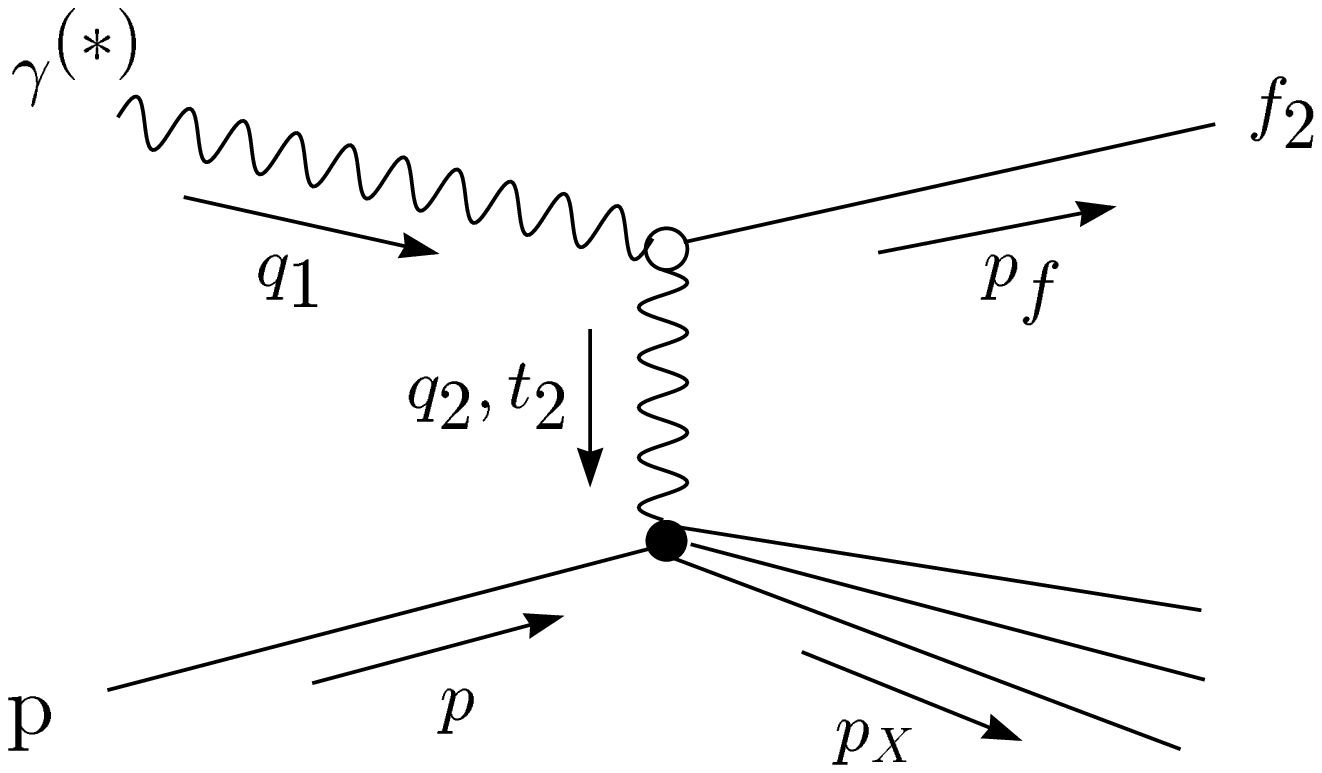}}

      \put(8.7,0){(b)}

      \put(3.3,2.8){$\mathbb{O}$}
      \put(11.6,2.8){$\gamma$}

      \put(5.9,1.3){X}
      \put(14.1,1.3){X}

    \end{picture}
  \end{center}
  \vspace*{-0.0cm}
  \caption{Feynman diagrams for  $f_2$ production in
the reaction  $\gamma^{( \ast )}{\rm{p}} \to f_2 {\rm{X}} $ 
at high energies with odderon (a) and photon (b) exchange.}
\label{b1}
\end{figure}

At sufficiently high energies only odderon and photon exchange contribute 
(Figure \ref{b1}) to these reactions. Pomeron exchange does not contribute
due to the positive charge parity of the $f_2$ (and of the pseudoscalars 
$\pi^0$ and $\eta^0$). Thus the energies available at HERA are an obvious 
attraction, and the cross sections, although small, are not unattainable.
The two reactions, photoproduction of the tensor meson $f_2(1270)$
and photoproduction of the pseudoscalars $\pi^0$ or $\eta^0$
complement each other from the experimental viewpoint as the problems
of detection and acceptance are very different. Given the rather
small cross sections, if a signal is found in one reaction it is
important to check that it is also there, at the appropriate level,
in another.

In Section 2 we outline briefly the formalism for tensor meson production 
in $\gamma^*$ p collisions. The next step is to construct a suitably
normalised wave function for the $f_2$ which is done in Section 3. The 
odderon-exchange contribution is calculated in Section 4 and the photon 
exchange contribution in Section 5. 
The latter turns out to be about
a factor of ten smaller than the odderon exchange.
The results are discussed in Section 6 and compared with 
what is currently known experimentally about $f_2$ photoproduction. We 
conjecture that the current data can be interpreted as providing some 
evidence for odderon exchange although alternative and less exciting 
interpretations cannot be excluded. Appendices A and B contain respectively 
technical details of the calculation of the $f_2$ wave functions and of 
their overlap with the photon wave function.
%
%
%
%%%%%%%%%%%%%%%%%%%%%%%%%%%%%%%%% FIGURE
%%%%%%%%%%%%%%%%%%%%%%%%%%%%%%%%%%%%
%
%
\newpage

\section{Tensor meson production in the $\gamma^{(*)}{\rm p}$ reaction}

We consider the production of the $J^{PC}=2^{++}$ 
tensor meson $f_2(1270)$ in the reaction 
\beqa
  \gamma^{\ast}(q_1) + {\rm p}(p) \rightarrow  
  f_2(p_f) + {\rm X}(p_X).
\label{reaction}
\eeqa
Here we treat this reaction  
for photon virtualities $Q^2 \le 5 \; {\rm{GeV}}^2$
where $q_1^2=-Q^2$ and in addition $q_2 = q_1-p_f$,
$t_2 = q_2^2$ and $W^2=s_2=(q_1+p)^2$.

We start with the
odderon exchange contribution of Figure \ref{b1}a, which we   
calculate in the path integral approach \cite{na91,dfk,all}. 
It was shown in \cite{etal} that the final state X is dominated
by the two negative parity isospin 1/2 resonances $N(1520)$ and
$N(1535)$ with spin 3/2 and 1/2 respectively. 

\begin{figure}[t]
 
\vspace*{0.0cm}
 
\hspace{1cm}
\epsfysize=4.0cm
\centerline{\epsffile{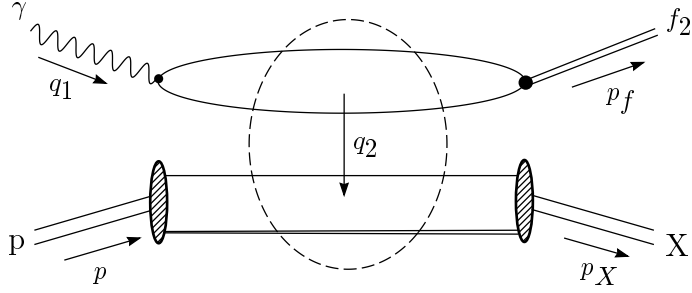}}
 
\vspace*{-0.5cm}
 
\caption{$f_2$ production in 
the reaction  $\gamma^{( \ast )}{\rm{p}} \to f_2 {\rm{X}} $. 
The dashed circle indicates the nonperturbative interaction
(odderon exchange)
of the colour dipoles.} 

\label{gampbild}
\end{figure}

In the following we are interested in unpolarised cross sections and 
sum over these two final state resonances. In this case we can neglect
the spin degree of freedom of the quark in the proton and the nucleonic
excitations, which we treat as quark diquark-states with 
a scalar diquark \cite{etal}. 
In this way we can describe the odderon-proton interaction as the
excitation of a spinless 
S-wave state (the proton) into a 2P state (either of the $N^*$ states).
The helicity amplitudes in the path integral
approach are given by:
\begin{eqnarray}
 &&T(s_2,t_2)_{\lambda,\lambda_f,\lambda_{\gamma}} = 
 2is_2 \int \, d^2b\,e^{i\vec{q_2}_T \vec{b}}\,
 \hat{J}_{\lambda,\lambda_f,\lambda_{\gamma}} (\vec b).
\label{helampl}
\end{eqnarray}
Here $\lambda$, $\lambda_f$ and $\lambda_{\gamma}$ are respectively the 
2P-state, the $f_2$
and the photon helicities. The profile function 
$\hat{J}$ is defined as
\begin{eqnarray}
 && \hat{J}(\vec{b})_{\lambda,\lambda_f,\lambda_{\gamma}} = 
 -\int \frac{d^2 r_1}{4\pi} dz
 \int \frac{d^2 r_2}{4\pi} 
 \sum_{q,h, \bar{h}} 
 \Psi^{* f}_{\lambda_f,\, q h_1 h_2}(\vec{r}_1,z)
 \Psi^{\gamma}_{\lambda_{\gamma},\,q h \bar{h}} (\vec{r}_1,z)
 \nonumber\\
 &&\hphantom{\hat{J}(\vec{b})_{\lambda,\lambda_f,\lambda_{\gamma}} =}
 \times
 \Psi^{*\, {\rm 2P}}_{\lambda} (\vec{r}_2)
 \Psi^{\rm p}(\vec{r}_2) \; \;
 \tilde{J}(\vec b, \vec{r}_1,z, \vec{r}_2).
\label{profile}
\end{eqnarray}
where
$z$ is the momentum fraction of the photon carried by the quark,
$q$ is a flavour index explained below (\ref{fnorm})
and 
$h,\bar{h}$ are respectively the quark and antiquark helicities.
The physical picture underlying (\ref{helampl}), (\ref{profile})
is shown schematically in Figure \ref{gampbild}.
The photon fluctuates into a $q\bar{q}$ pair, described by $\Psi^{\gamma}$.
By soft colour interaction (odderon exchange), calculated 
from the functional integral of two lightlike Wegner-Wilson loops
($\tilde{J}$), 
the $q \bar{q}$ pair turns into the tensor meson $f_2$ and the proton
($\Psi^{\rm p}$) is
excited into a 2P wave ($\Psi^{\rm 2P}$). An explicit expression for  
$\tilde{J}$ can be found in \cite{donaru}.

%
%
%
%%%%%%%%%%%%%%%%%%%%%%%%%%%%%%%%% FIGURE
%%%%%%%%%%%%%%%%%%%%%%%%%%%%%%%%%%%%%
%
%
%
From (\ref{helampl}) the cross section for $W^2>>m_{\rm{p}}^2$ is given by:
\begin{eqnarray}
  d^2 \sigma^{\mathbb{O}} = 
  \frac{1}{8 s_2} \frac{1}{(2 \pi)^2} d^2 k_T
  \sum_{\lambda}  
  \sum_{\lambda_f} 
  \sum_{\lambda_{\gamma}} 
  | T_{\lambda,\lambda_f,\lambda_{\gamma}} |^2.
  \label{cross}
\end{eqnarray}
The wave functions occuring in (\ref{helampl}) are light-cone wave functions
\cite{lightcone}. We use the photon wave 
function for low virtualities as derived in \cite{dogupi}
where it is argued that replacing the current quark mass 
in the expression for $\Psi^{\gamma}$ calculated in light cone 
perturbation theory 
by a $Q^2$-dependent constituent quark mass
$m(Q^2)$, with $m(0) \sim 0.21$ GeV,  leads to a wave function which can 
also be used for photon
virtualities $Q^2 \le 1$ GeV$^2$. The nucleon wave functions $\Psi^{\rm p}$
and $\Psi^{\rm 2P}$ can be found in \cite{etal}.

The model as it stands gives no energy dependence of the cross sections,
therefore all our numerical results refer to $W \approx 20$ GeV 
as the parameters are adapted to that energy.
We expect the energy dependence of the soft
odderon contribution  to be similar to that of the soft
pomeron as discussed in \cite{etal}. 
It should be recalled
that the results for photoproduction of mesons by odderon exchange are
particularly sensitive to the model parameters and the specific choice of the
photon and meson wave function. As explained in \cite{etal} we estimate
the overall
uncertainty in our results to be about a factor of two.

\section{The tensor meson wave function}

To construct a light cone wave function for the tensor meson $f_2$, 
we use a similar procedure as developed in \cite{guku} to construct 
a wavefunction for the $\rho$. 

We consider the $f_2$ to be a $q\bar{q}$ state consisting of on-shell quarks
of mass $m(Q^2)$, with $m(0) \sim 0.21$ GeV,
which moves with very large momentum in the 3-direction: 
$p_f = (p_f^0, \vec{0}_T, p_f^3)$ and $p_f^3 \rightarrow \infty$. 
To obtain the helicity structure of the wave function, we consider 
as interpolating operator the spin-two quark operator $O$,
\begin{eqnarray}
O^{\mu \nu}(x) :=  \frac{1}{2} \bar{\psi}(x) 
\Big\{
\gamma^{ \{ \mu} i 
{\buildrel \leftrightarrow \over \partial} {}^{\nu \}} -
{\textstyle \frac{1}{2}} i {\buildrel \leftrightarrow \over 
{\partial\hskip-6.6pt /\hskip2pt}} \, g^{\mu \nu} \Big\}
\,  \psi(x).
  \label{interpol}
\end{eqnarray}
For clarity we explain our derivation of the helicity structure for only
one flavour. The final result for the $f_2$ state will of course be written 
down with the appropriate flavour content.
The vertex factor  corresponding to (\ref{interpol}) is
\begin{eqnarray}
  &&V^{\mu \nu}_{h\bar{h}}(p_q,p_{\bar{q}}) := 
  \langle q(p_q,h), \, \bar{q}(p_{\bar{q}},\bar{h}) |
  O^{\mu \nu}(0) | 0 \rangle \nonumber\\
  &&\hphantom{V^{\mu \nu}_{h,\bar{h}}(p_q,p_{\bar{q}}) :} = 
  {\textstyle \frac{1}{2}} \bar{u}(p_q,h) 
  \Big\{ \gamma^{ \{ \mu } p_q^{\nu \}} - 
       \gamma^{ \{ \mu } p_{\bar{q}}^{\nu \} } 
  -m(Q^2) g^{\mu \nu}
  \Big\} v(p_{\bar{q}},\bar{h}).
  \label{vertex}
\end{eqnarray}
The quark and the antiquark four-vectors 
$p_q$ and $p_{\bar{q}}$ are parametrised by the relative transverse momentum
$\vec{k}_T$ and the momentum fraction $z$ of the $f_2$, carried by the 
quark. We have $p_q=(p_q^0, \vec{k}_T,z p_f^3)$, 
$p_{\bar{q}}=(p_{\bar{q}}^0, 
-\vec{k}_T,\bar{z} p_f^3)$ and $p_q^2=p_{\bar{q}}^2=m^2(Q^2)$
where $\bar{z}=(1-z)$.

Next we derive the polarisation tensors $e^{\mu \nu}(\lambda_f)$
for a massive spin-two particle. They can be constructed 
out of the polarisation vectors of a massive vector state
\begin{eqnarray}
&&e_{\pm}^{\mu}= (0,\mp 1,-i,0) / \sqrt{2},
\nonumber\\
&&e_0^{\mu}= (p_f^3,0,0,p_f^0)/m_f,
  \label{polvec}
\end{eqnarray}
by use of the appropriate Clebsch-Gordan coefficients \cite{particle} and 
are given by
\begin{eqnarray}
&&e^{\mu \nu} (\pm 2) = e_{\pm}^{\mu} e_{\pm}^{\nu},
\nonumber\\
&&e^{\mu \nu} (\pm 1) =
{\textstyle  \sqrt{\frac{1}{2}}}
(e_{\pm}^{\mu} e_0^{\nu} +  e_0^{\mu} e_{\pm}^{\nu}),
\nonumber\\
&&e^{\mu \nu} (0) =  
{\textstyle{  \sqrt{\frac{1}{6}}}}
(e_{+}^{\mu} e_-^{\nu} + e_{-}^{\mu} e_+^{\nu}) +
{\textstyle{  \sqrt{\frac{2}{3}}}} e_0^{\mu} e_0^{\nu}.  
  \label{polten}
\end{eqnarray}
Here we do not use directly the polarisation tensors (\ref{polten}) 
but replace the longitudinal polarisation vector $e_0$  in
(\ref{polten}) by $\tilde{e}_0 := e_0 - p_f/m_f$. This is justified
neglecting terms of order $m(Q^2)/m_f \le 15\%$. In any case this affects
only the helicity states $0,\pm 1$ of the $f_2$ which will be shown below 
to give only a small contribution to the cross section. 

Then we define the helicity structure $\tilde{\Psi}(\lambda_f)$,
for a given $f_2$ helicity $\lambda_f$ and quark and antiquark
helicity $h$ and $\bar{h}$ respectively, as the contraction 
of the vertex factor (\ref{vertex}) with the corresponding polarisation 
tensor $e^{\mu \nu}(\lambda_f)$,
\begin{eqnarray}
  \tilde{\Psi}_{h\bar{h}}(\lambda_f) := \lim_{p_f^3 \rightarrow \infty}
  \; \frac{1}{p_f^3} \, \sqrt{p_q^3 p_{\bar{q}}^3} \,
  e_{\mu \nu}(\lambda_f) V^{\mu \nu}_{h \bar{h}}.
 \label{heli}
\end{eqnarray}
%
%
%
%
%
%
%%%%%%%%%%%%%%%%%%%%%%%%%%%%%%%%% FIGURE
\begin{figure}[htb]
  \unitlength1.0cm
  \begin{center}
    \begin{picture}(15.,8.8)

      \put(0.4,1.0){
        \epsfysize=6.2cm
        \epsffile{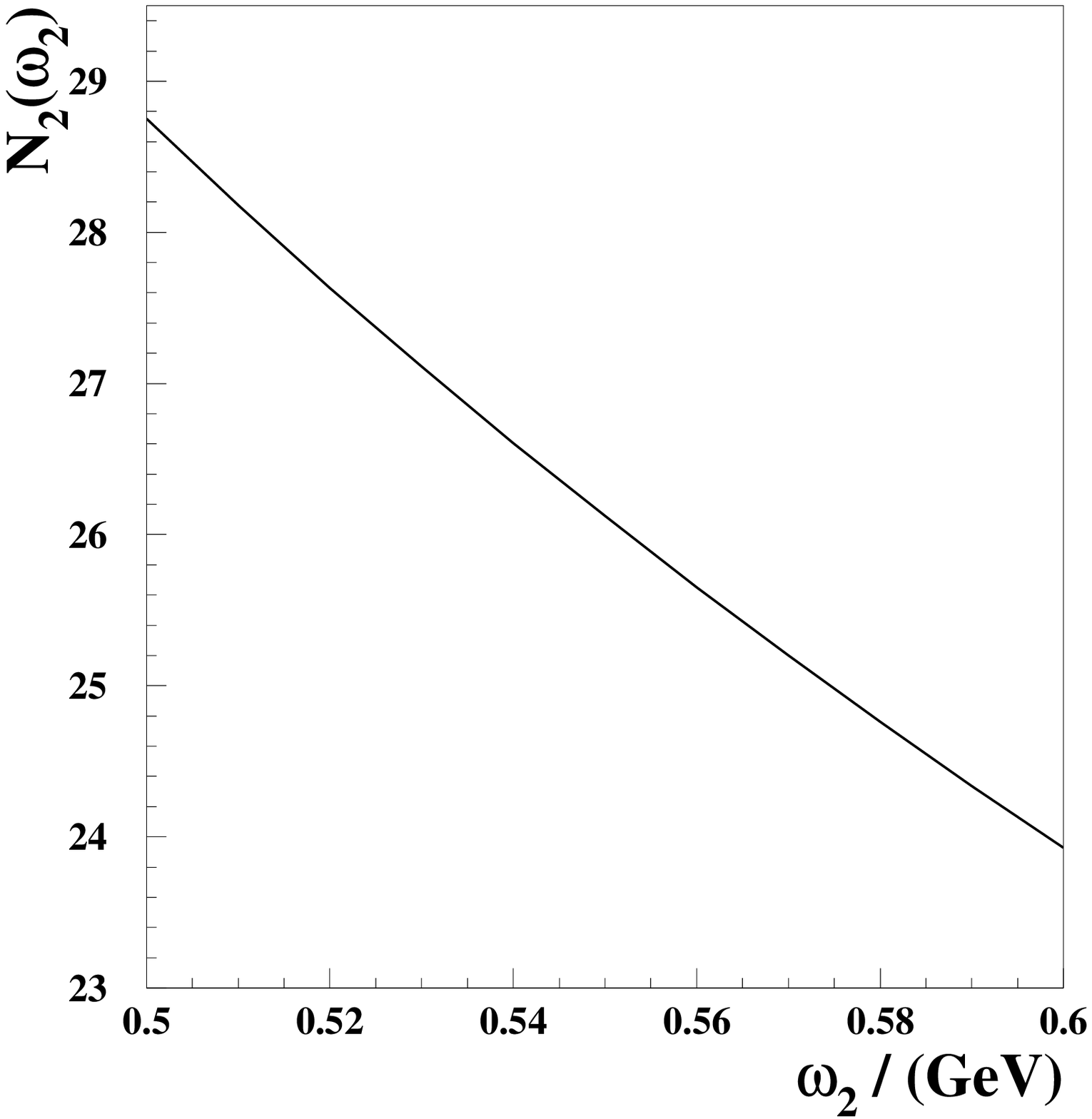}}

      \put(8.3,1.0){
        \epsfysize=6.2cm
        \epsffile{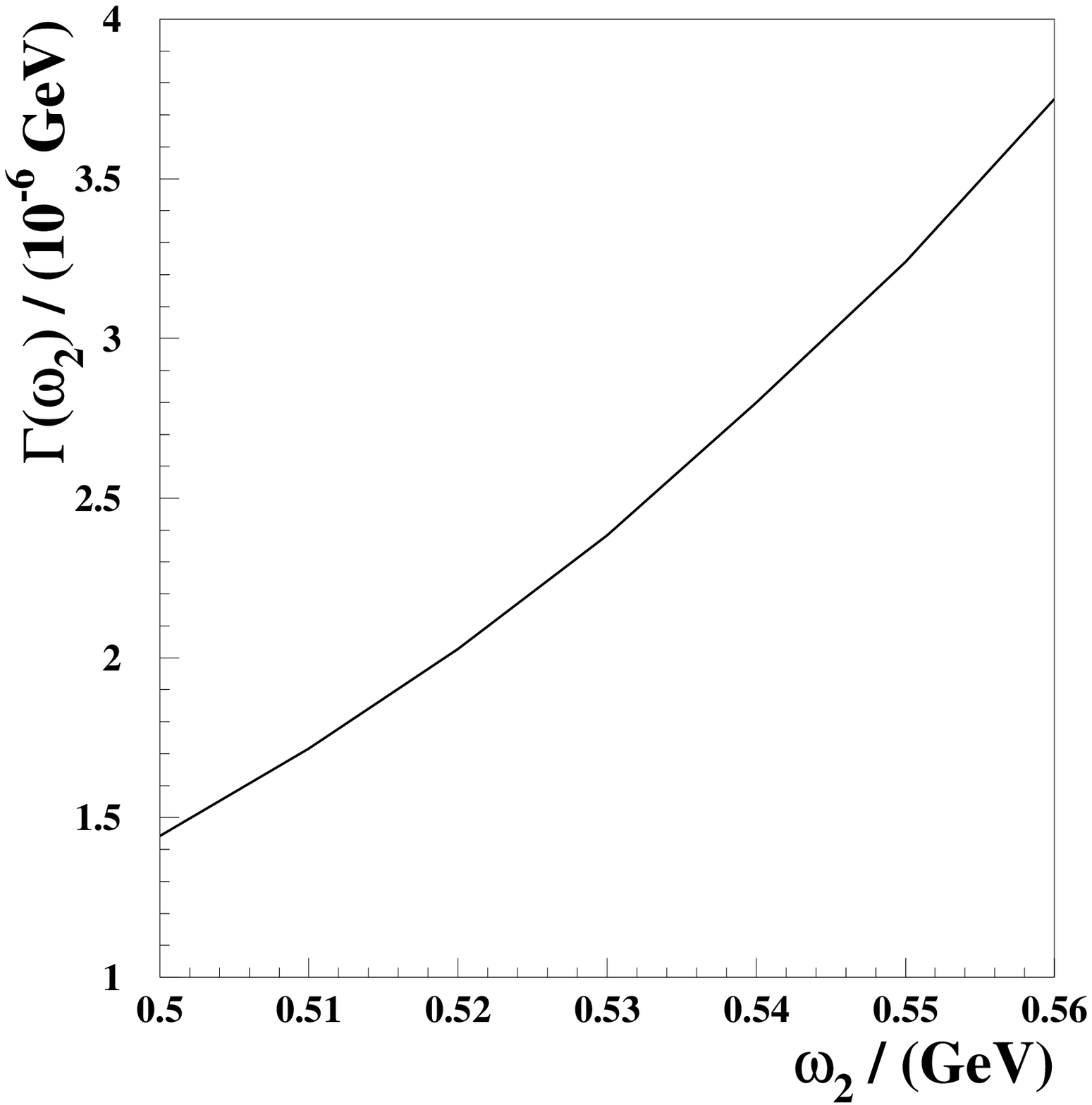}}

    \end{picture}
  \end{center}
  \vspace*{-1.4cm}
  \caption{(a) The normalisation constant $N_{2}$ 
as a function of $\omega_{2}$. (b) The partial decay width 
for a tensor meson $f_2$ with helicity 2 decaying into two photons
(\ref{decayres}) 
as a function of $\omega_{2}$.}
\label{omega}
\end{figure}
%%%%%%%%%%%%%%%%%%%%%%%%%%%%%%%%%%%%
%
%
The $f_2$ helicity wave functions are 
then given by multiplying 
$\tilde{\Psi}_{h\bar{h}}(\lambda_f)$ with  a Bauer-Stech-Wirbel (BSW) ansatz 
\cite{bsw} for 
the transverse and longitudinal momentum distribution of the quarks 
in the $f_2$,
\begin{eqnarray}
  \tilde{f}_{\lambda_f}(k_T,z) = 
  N_{\lambda_f} \, (z \bar{z})^{3/2} \,
  e^{-\frac{1}{2} m_f^2 (z-1/2)^2 / \omega_{\lambda_f}^2} \times 
  \frac{2 \pi}{\omega_{\lambda_f}^2 m_f} 
  e^{-\frac{1}{2} k_T^2/\omega_{\lambda_f}^2}.
  \label{bsw}
\end{eqnarray}
To define the helicity wave functions 
properly we first write down the $f_2$ state 
for helicity $\lambda_f$ as
follows:
\begin{eqnarray}
&&|f_2(p_f,\lambda_f) \rangle := \int_0^1 dz \int_{-\infty}^{+\infty}
\frac{d^2 k_T}{16 \pi^3} \frac{1}{\sqrt{z\bar{z}}} \; 
\sum_{q h \bar{h}}
\tilde{\Psi}^f_{\lambda_f,q h \bar{h}} 
\nonumber\\
&&\hphantom{|f_2(p_)f,\lambda_f \rangle := }
 \times 
{\textstyle{  \sqrt{\frac{1}{3}}}}
\delta_{A\bar{A}}
|q(p_q,h,A), \bar{q}(p_{\bar{q}},\bar{h},\bar{A}) \rangle
  \label{fstate}
\end{eqnarray}
with the normalisation  of the $f_2$ state chosen as 
\begin{eqnarray}
\langle f_2(p_f^{\prime},\lambda_f^{\prime} |
f_2(p_f,\lambda_f \rangle = 
2 p_f^0 (2 \pi)^3 \delta^3 (\vec{p}_f^{\prime}-\vec{p}_f)
\delta_{\lambda_f^{\prime}\lambda_f}.
  \label{fnorm}
\end{eqnarray}
The index $q$ in (\ref{fstate}) is a flavour index,
$q=u,d$, and $A,\bar{A}$ are colour indices.  
The wave functions $\tilde{\Psi}_{\lambda_f,qh\bar{h}}$ are  
the momentum 
space helicity wave functions of a quark-antiquark dipole 
with flavour $q$. They  follow directly from (\ref{heli}),(\ref{bsw})
and are given by
\pagebreak
\begin{eqnarray}
&& \tilde{\Psi}^f_{\pm 2,q h \bar{h}} = 
 (\pm 2) \Big\{ 
m(Q^2) \, (k_1 \pm ik_2) \delta_{h \pm } \delta_{\bar{h}\pm} +
\nonumber\\
&&\hphantom{\tilde{\Psi}^f_{\pm 2,q h \bar{h}} = 
 (\pm 2) \Big\{ } 
(k_1\pm i k_2)^2 
\big( z \delta_{h \pm } \delta_{\bar{h}\mp} - 
     \bar{z} \delta_{h \mp } \delta_{\bar{h}\pm} \big) \, \Big\}
c_q \; \tilde{f}_2(k_T,z),
\nonumber\\
&&  \tilde{\Psi}^f_{\pm 1,q h \bar{h}} = 
(-1) \Big\{
m_f \, m(Q^2) \, (z-\bar{z}) \delta_{h \pm} \delta_{\bar{h} \pm} \mp
\nonumber\\
&&\hphantom{\Psi^f_{+1,q h \bar{h}} = \frac{1}{2}(-) \Big\{}
m_f \, (k_1\pm ik_2) \Big(
(3z-4z^2)\delta_{h \pm } \delta_{\bar{h}\mp } +
(3\bar{z}-4\bar{z}^2) \delta_{h\mp } \delta_{\bar{h}\pm } \Big) \Big\}
c_q \; \tilde{f}_1(k_T,z),
\nonumber\\
&&  \tilde{\Psi}^f_{0,q h \bar{h}} = 
(-
{\textstyle{  \sqrt{\frac{2}{3}}}}
) \Big\{
m(Q^2) \,  \Big( (k_1-ik_2)\delta_{h+} \delta_{\bar{h}+} -
(k_1+ik_2)\delta_{h-} \delta_{\bar{h}-}  \Big) +
\nonumber\\
&&\hphantom{\Psi^f_{0,q h \bar{h}} = ({\textstyle -\sqrt{\frac{2}{3}}}) 
\Big\{}(k_T^2 + 2 z \bar{z} m_f^2 )
\Big( (z-\bar{z}) \delta_{h+} \delta_{\bar{h}-} -
(\bar{z}-z) \delta_{h-} \delta_{\bar{h}+} \Big) \Big\}
c_q \; \tilde{f}_0(k_T,z).
  \label{kresult}
\end{eqnarray}
where $c_q=1/\sqrt{2}$ for $q=u,d$.
The constants $N_{\lambda_f}$ 
are determined for each helicity by  normalisation as a function of
the frequencies $\omega_{\lambda_f}$. 

From (\ref{fstate}),(\ref{fnorm})
it follows that the normalisation conditions of our helicity wave functions
(\ref{kresult}) are:
\begin{eqnarray}
\int_0^1 dz \int_{-\infty}^{+\infty}
\frac{d^2k_T}{16 \pi^3} \,
\sum_{q ,h , \bar{h}}
| \tilde{\Psi}^f_{\lambda_f,q h \bar{h}}(\vec{k}_T,z) |^2 = 1.
  \label{wavenorm}
\end{eqnarray}
In Figure \ref{omega}a we show $N_{2}$ 
as a function of $\omega_{2}$. 
To fix the frequencies $\omega_{\lambda_f}$
we need additional conditions. We use the partial decay width of the
$f_2$ decaying into two photons, $\Gamma_{f_2 \rightarrow \gamma \gamma}$,
with the central value \cite{particle} being $\Gamma_{f_2 \rightarrow
\gamma \gamma} = 2.4$ keV. From
rotational invariance it follows that the $f_2$ decays with the same decay
width into two photons, independent of its helicity,
\begin{eqnarray}
\Gamma(f_2(\lambda_f) \rightarrow \gamma \gamma) = 
\Gamma_{f_2\rightarrow \gamma \gamma}.
\label{equaldecay}
\end{eqnarray}
In Appendix A we consider the $f_2$-decay for each helicity 
in the infinite momentum frame and
calculate $\Gamma(f_2(\lambda_f) \rightarrow \gamma \gamma)$,
defined in (\ref{decay}),
as a function of the corresponding $\omega_{\lambda_f}$. 
In Figure \ref{omega}b we show our result, again for the case $\lambda_f=2$.
Then $\omega_{2}$ is fixed by requiring (\ref{equaldecay}). Using the
value of $\omega_{2}$ thus obtained we can read off from 
Figure \ref{omega}a the corresponding normalisation factor
$N_2$. For the other helicities we proceed in a
similar way. The results for the normalisation constants and frequencies
are listed in Table \ref{normres}
($\omega_{\lambda_f}=\omega_{-\lambda_f},
N_{\lambda_f}=N_{-\lambda_f}$).
\begin{table}[t]
\begin{center}
$
\begin{array}{|l||c|c|c|}
  \hline
  & \lambda = 2  
  & \lambda = 1  
  & \lambda = 0  \\
 \hline
  N_{\lambda_f} & 27.11 & 38.78 & 47.33 \\
  \hline
  \omega_{\lambda_f}    & 0.53 & 0.60 & 0.36 \\
  \hline
\end{array}
$
\caption{The numerical values for the  
         normalisation constant $N_{\lambda_f}$
and for the frequencies $\omega_{\lambda_f}$.}
\label{normres}
\end{center}
\end{table}
Now the momentum space wave functions (\ref{kresult}) are fixed
and it remains only to calculate the Fourier-transform in order to get  
the configuration-space wave functions needed in (\ref{helampl}).
This is done in Appendix B.

\section{The odderon contribution}

Now we come back to the helicity amplitudes. To calculate
(\ref{helampl}) we first have to consider the overlap 
functions between the photon and the
$f_2$ wave functions. The explicit results  are listed in
(\ref{transov1}), (\ref{longov}), (\ref{transov2}) of Appendix B.
As we can see there the dependence of all overlap functions
on $\theta_1$, the angle between
$\vec{q}_{2_T}$ and $\vec{r}_1$, is 
given by a phase,
\begin{eqnarray}
  &&\sum_{f,h,\bar{h}} 
  \Psi^{\ast f}_{\lambda_f,qh\bar{h}}
  \Psi^{\gamma}_{\lambda_{\gamma},qh\bar{h}} =
  e^{i (\lambda_{\gamma}-\lambda_f)\theta_1} \,
  \Big[
  \sum_{f,h,\bar{h}} 
  \Psi^{\ast f}_{\lambda_f,qh\bar{h}}
  \Psi^{\gamma}_{\lambda_{\gamma},qh\bar{h}}
  \Big]_{\theta_1 = 0}.
  \label{zero}
\end{eqnarray}
Inserting (\ref{zero}) in (\ref{helampl}) we can integrate over the 
angle $\theta_b$,
which is the
angle between $\vec{b}_T$ and $\vec{q}_{2_T}$. We
choose as new integration variables in (\ref{helampl}) the relative 
angles between $\vec{b}_T$ and $\vec{r}_{1(2)}$, 
$\theta_{1(2)}^{\prime} = \theta_{1(2)} - \theta_b$, where $\theta_2$ 
is the angle between $\vec{q}_{2_T}$ and $\vec{r}_2$. 
In this way 
$\tilde{J}$ in (\ref{helampl}) becomes independent of $\theta_b$.
By performing similar steps leading to (12) in \cite{etal} we get 
as the result for the helicity amplitudes:
\begin{eqnarray}
 &&T_{\lambda,\lambda_f,\lambda_{\gamma}} = 2 i s_2 \int b\, db\,
 \int \frac{d^2 r_1}{4\pi} dz
 \int \frac{d^2 r_2}{4\pi}
 \nonumber\\
&&\hphantom{T_{\lambda_{\gamma}}}
 \times 
 \Big[ \sum_{q,h, \bar{h}}
 \Psi^{* f}_{\lambda_f,q h \bar{h}}(r_1,z)
 \Psi^{\gamma}_{\lambda_{\gamma},q h \bar{h}}(r_1,z) \Big]_{\theta_1=0}
 \Psi^{*\, 2{\rm P}}(r_2)\Psi^{\rm p}(r_2)
 \nonumber\\
&&\hphantom{T_{\lambda_{\gamma}}}
 \times 
 e^{i ((
\lambda_{\gamma}-\lambda_f) \theta_1^{\prime} +
\lambda \theta_2^{\prime})}
 (i)^{(\lambda +\lambda_{\gamma}-\lambda_f)} 
 2\pi 
 J_{(\lambda +\lambda_{\gamma}-\lambda_f)}(\sqrt{-t_2}b) \; 
 \tilde{J}(\vec{b}_T, \vec{r}_1,z, \vec{r}_2).
 \label{gampresult}
\end{eqnarray}
We note that for $\lambda = \lambda_f-\lambda_{\gamma}$ the forward
amplitude does not vanish.

\begin{figure}[htb]
 
  \vspace*{-1.0cm}
  
  \hspace{1cm}
  \epsfysize=6.5cm
  \centerline{\epsffile{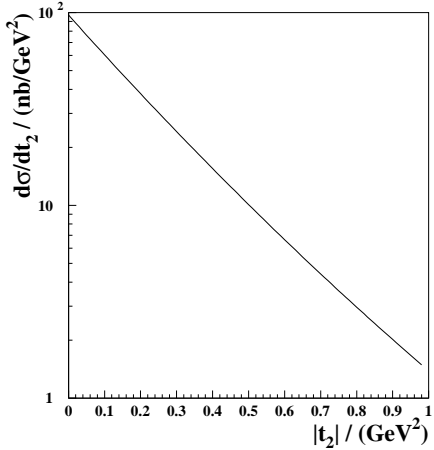}}
  
  \vspace*{0cm}
  
  \caption{The differential cross section for 
    $\gamma {\rm{p}} \rightarrow f_2 {\rm{X}}$. 
    A fit to this curve is:
    $d \sigma_{t}^{\mathbb{O}}/d t_2= a \,
     \exp(-b|t_2| + c |t_2|^2)$ where
     $a=97 \; {\rm nb/GeV}^2, b=4.8 \; {\rm GeV}^{-2},
    c=0.52 \; {\rm GeV}^{-4}$.}
\label{dsigoddt}
\end{figure}

We first consider $f_2$ production by real photons. 
Using an argument similar to that in \cite{etal} we see that the 
amplitudes for $\lambda = 0 $ vanish, so
we have 20 helicity amplitudes contributing to the sum in
(\ref{cross}). However, from (\ref{transov1}) and (\ref{transov2}) and 
noting that $J_{-n}(x)=(-1)^n J_n(x)$ it follows that
\begin{eqnarray}
  |T_{\lambda,\lambda_f,\lambda_{\gamma}}|^2 = 
  |T_{-\lambda,-\lambda_f,-\lambda_{\gamma}}|^2
 \label{mi10}
\end{eqnarray}
and we are left with only 
10 independent amplitudes. Of course (\ref{mi10}) is a consequence of parity
invariance.
The result for the
differential cross section is:
\begin{eqnarray}
  \frac{d \sigma_{t}^{\mathbb{O}}}{d t_2} = 
  \frac{1}{16 \pi} \frac{1}{s_2^2} \,
  \sum_{\lambda_f} \sum_{\lambda_{\gamma}}
  | T_{1,\lambda_f,\lambda_{\gamma}} |^2
 \label{sigt}
\end{eqnarray}
We show our result for the differential
cross section in Figure \ref{dsigoddt}. 
%
%
%
%%%%%%%%%%%%%%%%%%%%%%%%%%%%%%%%% FIGURE
%%%%%%%%%%%%%%%%%%%%%%%%%%%%%%%%%%%%%
%
%
%
A first observation is that it looks quite similar to
the differential cross section of $\pi^0$ production \cite{etal}, but
the normalisation is more than a factor of ten smaller. This can
be understood by considering the flavour part  
of the $\pi^0$
and the $f_2$ wave functions. The $\pi^0$ has isospin one,
so ${\pi^0 \sim (u\bar{u}} - d\bar{d})/\sqrt{2}$. This leads to a
factor e/$\sqrt{2}$ in the $\pi^0 \gamma$-overlap as the 
flavour part of the photon wave function is
$\gamma \sim (2 u\bar{u} - d\bar{d})/3$. 
On the other hand in the case of the $f_2$ we have
${f_2 \sim (u\bar{u}} + d\bar{d})/\sqrt{2}$ and so 
we get 
$e/\sqrt{18}$ for the $\gamma f_2$ overlap. This can simply be restated that
in the case of the $\pi^0$ the photon must be isovector while in the case of
the $f_2$ the photon must be isoscalar.
These factors enter 
quadratically into the cross section,
so naively we expect the $f_2$ photoproduction cross section to be a 
factor of 9 smaller than that for the $\pi^0$.

\begin{figure}[htb]
  \unitlength1.0cm
  \begin{center}
    \begin{picture}(15.,8.8)

      \put(0.5,1.0){
        \epsfysize=6.5cm
        \epsffile{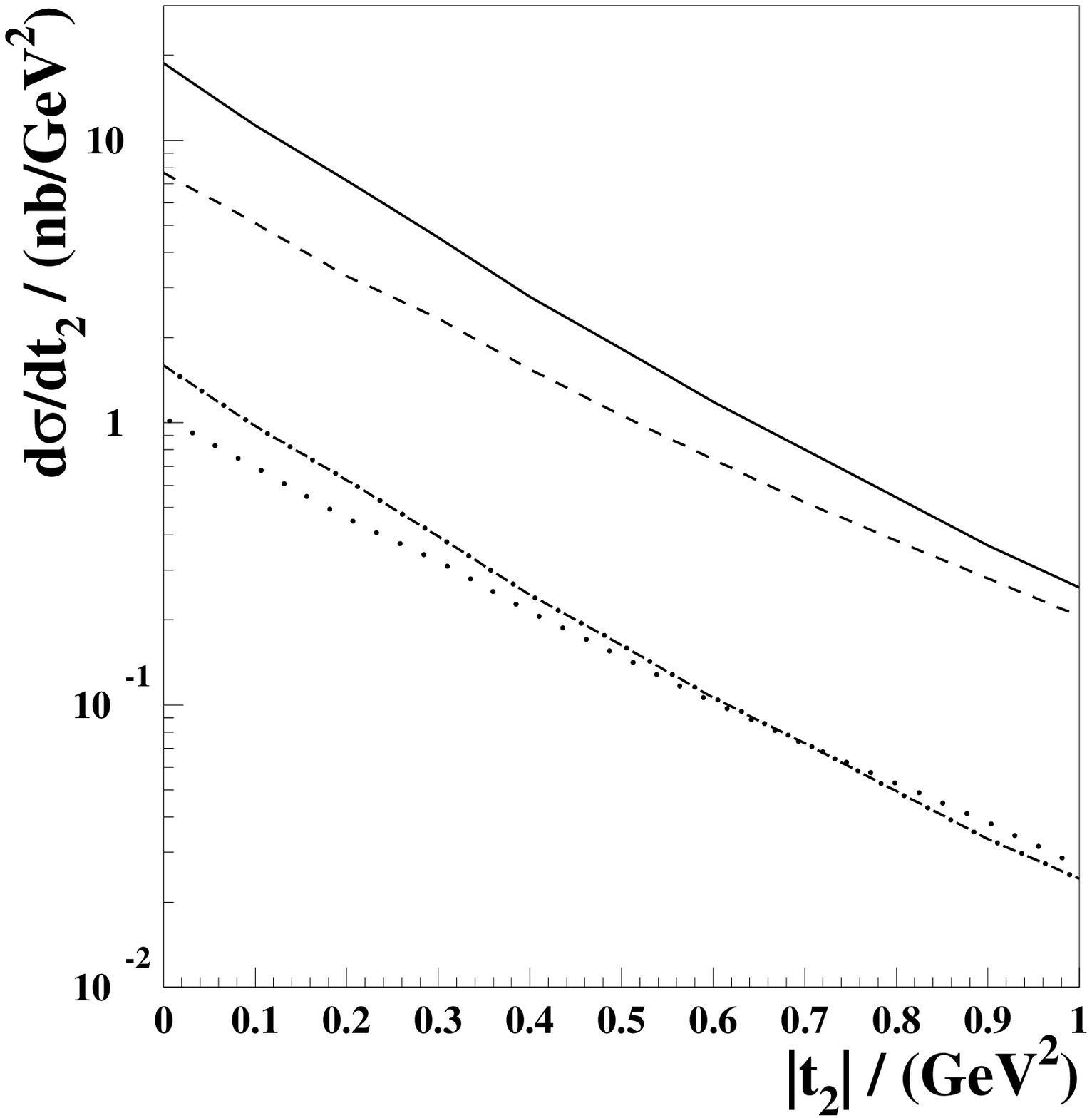}}

      \put(8.2,1.0){
        \epsfysize=6.5cm
        \epsffile{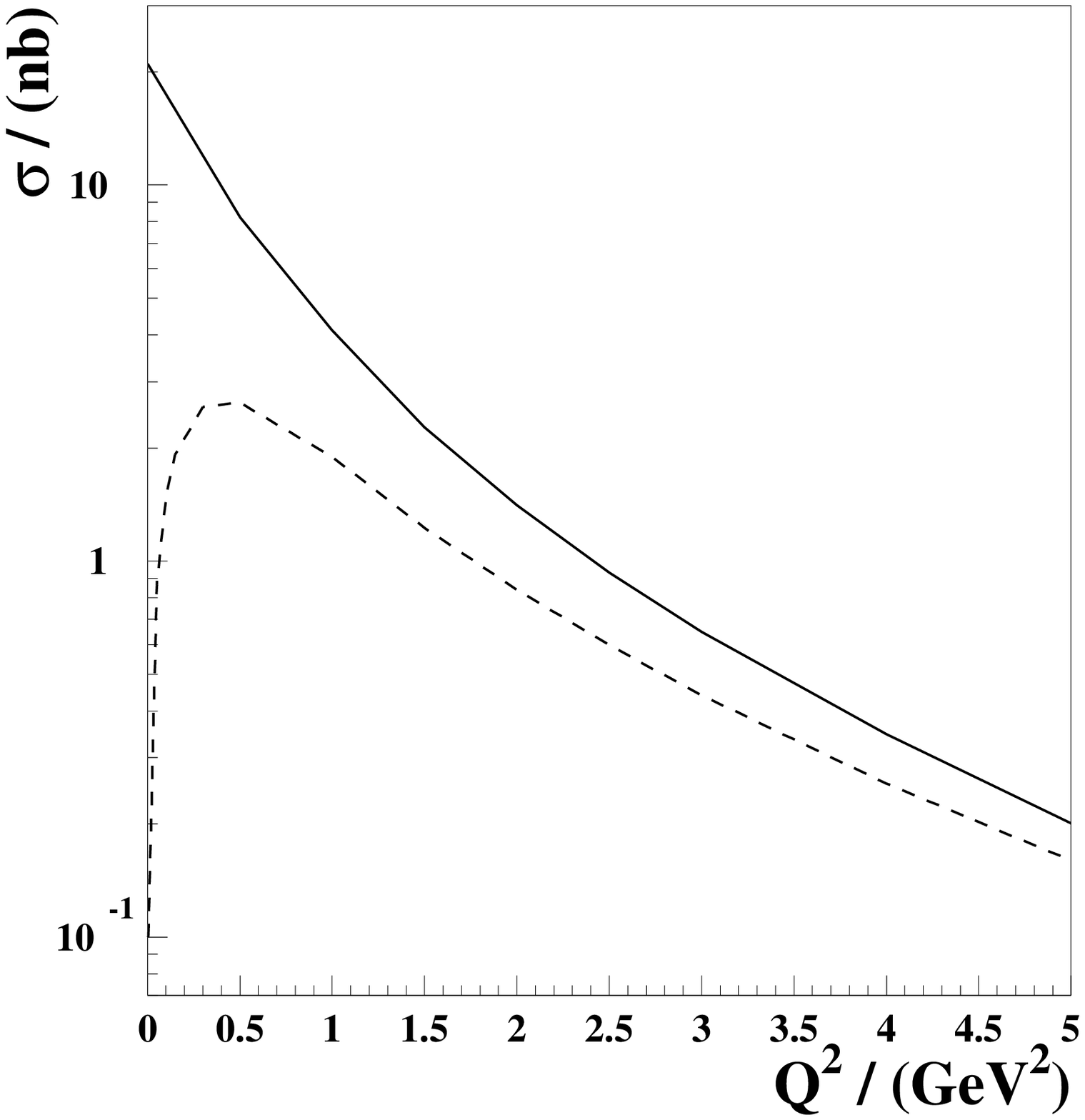}}

    \end{picture}
  \end{center}
  \vspace*{-1.3cm}
  \caption{(a) The differential cross section for 
  $\gamma^* {\rm{p}} \rightarrow f_2 X$
  for $Q^2 = 1\; {\rm{GeV}}^2$ for transversely polarised photons
  (solid line) and  longitudinally polarised photons (dashed line)
  and for
  $Q^2 = 4\; {\rm{GeV}}^2$ for 
  transversely polarised photons
  (dashed dotted line) 
  and for longitudinally polarised photons (dotted line). (b) The
  integrated cross section for transversely (solid line) and
  longitudinally (dashed line) polarised photons.}
\label{dsigodd}
\end{figure}

Second, by computing the various helicity amplitudes we find that
almost all of the cross section comes from the amplitudes $T_{1,2,1}$
and $T_{-1,-2,-1}$, where a helicity $\pm 1$ photon is diffractively
transformed into a helicity $\pm 2$ tensor meson $f_2$. 
This means that the conservation of $s$-channel helicity is almost
maximally violated at the particle level although it is fulfilled at the
quark level. 
By integrating the differential distribution of
Figure \ref{dsigoddt} we get for the total cross section:
\begin{eqnarray}
&&\sigma^{\mathbb{O}}
(\gamma \; {\rm p} \to f_2 \; 2{\rm P}) = 
21 \;{\rm nb}.
\label{difftot}
\end{eqnarray}
Next we calculate the differential cross section for $Q^2\neq 0$. In
addition to (\ref{sigt}) we define the longitudinal differential cross
section as
\begin{eqnarray}
  \frac{d \sigma_{l}^{\mathbb{O}}}{d t_2} = 
  \frac{1}{16 \pi} \frac{1}{s_2^2} \,
  \sum_{\lambda}
  \sum_{\lambda_f}
  | T_{\lambda,\lambda_f,0} |^2.
 \label{sigl}
\end{eqnarray}
In Figure \ref{dsigodd}a we compare 
the differential cross section for transversely and longitudinally
polarised photons colliding with the proton for virtualities
$Q^2 = 1,4 \; {\rm{GeV}}^2$. In addition we show in Figure
\ref{dsigodd}b the integrated transverse and longitudinal cross section
for $Q^2\le5\;{\rm GeV}^2$, which is calculated by integrating
(\ref{sigt}), (\ref{sigl}).  
%
%
%
%%%%%%%%%%%%%%%%%%%%%%%%%%%%%%%%% FIGURE
%%%%%%%%%%%%%%%%%%%%%%%%%%%%%%%%%%%%
%
%
%
With increasing photon virtuality the
transverse cross section drops quite rapidly. From $Q^2 =0 $ to 
$Q^2=1 \;{\rm{GeV}}^2$ 
the cross section decreases by more than  a factor of five. 
At $Q^2=4 \;{\rm{GeV}}^2$ the longitudinal cross section becomes
comparable to the transverse one. 
However the decrease with increasing $Q^2$ is not so rapid as in
the case of $\pi^0$ production by odderon exchange \cite{donaru}.
Due to the dependence of the constituent mass $m(Q^2)$ on $Q^2$
for $Q^2 \leq 1$ GeV$^2$, the cross section depends strongly on the 
weight of the terms in the overlap integrals which are proportional
to $m(Q^2)$. For $f_2$ production this contribution is rather small
(about $20\%$ at $Q^2 = 0$). For $\pi^0$ production the terms 
in $m(Q^2)$ are significant, resulting in a steeper fall-off with
increasing $Q^2$ as can be seen from Figure 8 of \cite{donaru}.

\begin{figure}[htb]
 
  \vspace*{-1.0cm}
  
  \hspace{1cm}
  \epsfysize=7.5cm
  \centerline{\epsffile{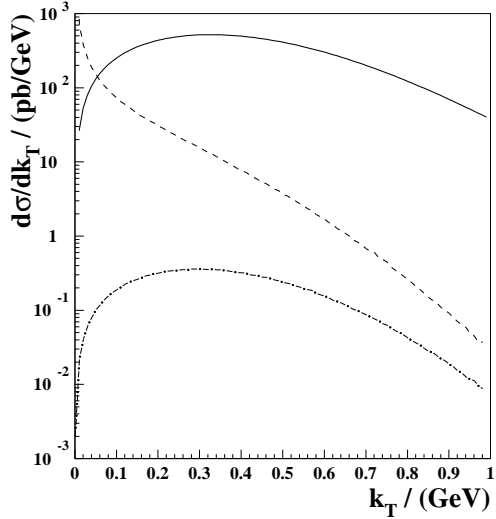}}
  
  \vspace*{0cm}
  
  \caption{Transverse momentum distribution 
   of the $f_2$ produced in ep collisions for odderon exchange (solid line),
   for photon exchange when the proton stays intact (dashed line) and for
   photon exchange, when the proton gets excited into the resonances 
   N(1520), N(1535) (dashed dotted line).}
\label{ktodd}
\end{figure}

Finally we go from $\gamma^*$p to ep, restricting 
ourselves
to small photon virtualities, $Q^2 < 0.01 \; {\rm{GeV}}^2$ and 
applying
the equivalent photon approximation (EPA) \cite{epa}. 
As the experimentally prefered observable we calculate the 
transverse
momentum ($k_T$) distribution of the tensor meson $f_2$ with 
respect to
the ep collision axis. Although $k_T$
is formally defined as the transverse momentum  of the $f_2$ 
with respect 
to the
incoming photon we are allowed to
identify it with the transverse momentum relative to the ep 
collision 
axis due to the restriction $Q^2 < 0.01 \; {\rm{GeV}}^2$, at
least for $k_T>0.1$ GeV.
\begin{eqnarray}
  \frac{d \sigma_{{\rm ep}}^{\mathbb{O}}}
  {d |\vec{k}_T|} = c_{\rm{EPA}}
  \frac{1}{8 \pi} \frac{1}{W^2} |\vec{k}_T| \,
  \sum_{\lambda_f} \sum_{\lambda_{\gamma}}
   | T_{1,\lambda_f,\lambda_{\gamma}} |^2.
  \label{epktodd}
\end{eqnarray}
Since the cross section (\ref{cross}) does not depend on 
the energy we
just have to multiply it with a constant $c_{\rm{EPA}}$
given by the integral over the
equivalent photon spectrum of the incoming electron. This constant  
depends on the
phase space cuts applied. For the HERA ``photoproduction'' 
cuts (which in addition to $Q^2 < 0.01 \; {\rm{GeV}}^2$ require
$0.3 \le y \le 0.7$ where $y$ is 
the energy fraction of the incoming electron carried by the
photon in the proton rest frame) $c_{\rm{EPA}}=0.0136$ \cite{etal}. 
The result for the  $k_T$ distribution is shown as the solid line
in Figure \ref{ktodd}.
This gives a total cross section for ${\rm e p} \rightarrow {\rm e}f_2{\rm 2P}$
by odderon exchange of $\sim 285$ pb.
%
%
%
%%%%%%%%%%%%%%%%%%%%%%%%%%%%%%%%% FIGURE
%%%%%%%%%%%%%%%%%%%%%%%%%%%%%%%%%%%%%
%
%
%
\section{The photon contribution}

In this section we calculate the electromagnetic contribution
Figure \ref{b1}b. 
In contrast to diffractive $f_2$ production, in the electromagnetic case 
we get a contribution from ``elastic''
production, where the proton stays intact, 
as well as from  ``inelastic''  production.
In the following we calculate the two cases where 
either the 
proton stays intact or the proton gets exited into 
the resonances N(1520), N(1535).

We start with the  ``inelastic'' case and calculate it in a 
similar way to the diffractive contribution. To do so we just have 
to do the replacement $\tilde{J} \rightarrow \tilde{J}_q^{\gamma}$
in (\ref{helampl}),(\ref{profile}) where to lowest order in the
electromagnetic coupling constant $\tilde{J}_q^{\gamma}$ describes one
photon exchange between the $q\bar{q}$ pair from the photon and the
quark-diquark pair of the proton,
\begin{eqnarray}
  &&\tilde{J}_q^{\gamma} = -i e^2 Q_q \bigg\{
  \frac{2}{3} 
  \frac{K_0(\mu |\vec{b} + z\vec{r}_{1} - \vec{r}_{2}/2|}{2 \pi} -
  \frac{2}{3} 
  \frac{K_0(\mu |\vec{b} - \bar{z} \vec{r}_{1} - \vec{r}_{2}/2|}{2 \pi} +
  \nonumber\\
  &&\hphantom{\tilde{J}_q^{\gamma} = -i e^2 Q_q \Big\{}
  \frac{1}{3} 
  \frac{K_0(\mu |\vec{b} + z \vec{r}_{1} + \vec{r}_{2}/2|}{2 \pi} -
  \frac{1}{3} 
  \frac{K_0(\mu |\vec{b} - \bar{z}\vec{r}_{1} + \vec{r}_{2}/2|)}{2 \pi}
  \bigg\} .
 \label{jphoton}
\end{eqnarray}
Here we have introduced a photon mass $\mu$ 
which will disappear in our final results. In addition $Q_q$ is the
quark charge,
$Q_u=2/3, Q_d = -1/3$. 

We use (\ref{helampl}),(\ref{profile}), 
replacing $\tilde{J}$ by $\tilde{J}_q^{\gamma}$. In this way 
each helicity amplitude consists of four terms proportional to the
photon propagators in configuration space. Making a shift $\vec{b}$ 
in an appropriate way we can perform the $\vec{b}$ integrations
in each of these contributions. In this way
the photon propagator in momentum space enters.
Then, finally we can also perform 
the $\theta_1,\theta_2$
integrations to  get as result for the ``inelastic''
electromagnetic part of the helicity amplitudes:
\begin{eqnarray}
  &&T_{\lambda,\lambda_f,\lambda_{\gamma}}^{\gamma} =
  2 i s_2 
  \int \frac{d r_1 r_1}{4 \pi} dz
  \int \frac{d r_2 r_2}{4 \pi}
  \Big[ \sum_{q,h, \bar{h}}
  \Psi^{*f}_{\lambda_f,q h \bar{h}}(r_1,z)
  \Psi^{\gamma}_{\lambda_{\gamma},q h \bar{h}}(r_1,z) \Big]_{\theta_1=0}
 \nonumber\\
  &&\hphantom{T_{\lambda,\lambda_f,\lambda_{\gamma}} =}\times
  \Psi^{*\, 2{\rm P}}(r_2)\Psi^{\rm p}(r_2)
  (\frac{5 i e^2}{9} ) \frac{1}{|t_2|} \, 
  (2 \pi)^2 \, i^{\lambda + \lambda_{\gamma} - \lambda_f}
 \nonumber\\
  &&\hphantom{T_{\lambda,\lambda_f,\lambda_{\gamma}} =}\times
  \Big( J_{\lambda_{\gamma}-\lambda_f} (\sqrt{|t_2|} z r_1) -
  (-1)^{\lambda_{\gamma}-\lambda_f}
  J_{\lambda_{\gamma}-\lambda_f} (\sqrt{|t_2|} \bar{z} r_1)\Big) 
 \nonumber\\[0.05cm]
  &&\hphantom{T_{\lambda,\lambda_f,\lambda_{\gamma}} =}\times
  \Big( 2             J_{\lambda} (\sqrt{|t_2|} r_2/2) +
   (-1)^{\lambda} J_{\lambda} (\sqrt{|t_2|} r_2/2) \Big).
 \label{helamplgam}
\end{eqnarray}
The result for the ``elastic'' helicity amplitudes 
$T_{\lambda_f,\lambda_{\gamma}}^{\gamma}$ can be read off from
(\ref{helamplgam}) for $\lambda=0$, replacing $\Psi^{\rm{2P}}$ by 
$\Psi^{\rm{p}}$.

We concentrate directly on the ep reaction and apply the EPA,
restricting ourselves again to HERA photoproduction cuts,
and calculate the $k_T$ spectrum
(\ref{epktodd})
for the ``elastic'' and the ``inelastic'' case. The result 
is shown in Fig. \ref{ktodd}. 
As in $\pi^0$ production photon exchange is dominated by odderon
exchange
for $k_T\ge  0.1$ GeV, but not by as much as in the former case. 
This can again be understood from consideration of
the flavour parts.
We couple two photons to the upper quark loop, which leads to a factor
$(Q_q e)^2$. Now when summing over the flavours $q=u,d$, taking into
account the flavour parts of the wave functions, we find in case of the
$\pi^0$ a factor $1/9$ and in case of the $f_2$ a factor $5/9$. 
However, already for $k_T \ge 0.2$ the odderon contribution is 
more than a factor of ten larger than the electromagnetic 
contribution. 

The integrated cross section gets most of its
contributions from very small values of $|t_2|$, so the result is 
sensitive to the lower limit of $|t_2|$. Here we are specially
interested in the value for the E687 experiment, where $\sqrt{s_2}$ 
is between 7 and 17 GeV. For the averaged value 
$\sqrt{s_2}=12$ GeV we get a lower limit of about 0.0001 
${\rm{GeV}}^2$. Using it we 
find for the ep reaction as integrated cross section
for the electromagnetic contribution:
\begin{eqnarray}
  \sigma_{e{\rm{p}}}^{\gamma} ({\rm{e \, p}} 
  \rightarrow {\rm{e}} \, f_2 \, {\rm{p}} ) \approx 30 \, {\rm{pb}}
\end{eqnarray}
which is about ten times smaller than the odderon cross section. 
As can be seen from Figure \ref{ktodd} 
the breakup by photon exchange is about a
factor of 1000 smaller than the one induced by odderon exchange; interference
terms between the odderon and the photon exchange are therefore to be expected
at most on the 10 percent level.

\section{Discussion}

In this paper we have calculated the diffractive photo- and electroproduction
of the $f_2$ at high energies. Our approach is based on functional integral 
techniques and the model of the stochastic vacuum to treat QCD in the
nonperturbative region. We find a cross section of $\sim 20$ nb for the
odderon-exchange reaction $\gamma{\rm p} \rightarrow f_2{\rm X}$ where the 
proton is required to break up. 
We estimate the overall uncertainty of our results to be about
a factor of 2.
In our model the elastic reaction 
$\gamma{\rm p} \rightarrow f_2{\rm p}$ gets no contribution from odderon
exchange in the strict quark-diquark limit for the proton structure.
We also calculated the contributions from photon exchange instead of the 
odderon and found them to be small. We turn now to a discussion of the 
relevant experimental information.

In their high-statistics study of diffractively produced $\pi^+\pi^-$
states at high energy, $\langle \surd{s} \rangle \sim 12$ GeV, the E687 
Collaboration \cite{E687} has shown unambiguous evidence for the presence 
of the $f_2(1270)$. Its strength is between $\sim 0.1\%$ and $\sim 0.16\%$ 
of the $\rho$ signal, depending on whether one or two $\rho'$ states
are included in the analysis. As an explanation of its observation the
E687 collaboration suggested that as some particles, for example neutrals,
could possibly be missed at such low relative yields the $f_2(1270)$
signal should be considerd as part of the background. However there are
at least two other explanations. One is the mundane one of Regge exchange,
specifically $\rho$ and $\omega$, as the energy of the experiment is not
sufficiently high to exclude this possibility. The second is, of course,
odderon exchange.

Cross sections are not quoted explicitly by E687 but production rates
relative to the $\rho$ are given. The experiment, which was designed 
primarily for charm photoproduction, used a 4cm long beryllium target
and selected the $\pi^+\pi^-$ candidates from two-prong events with a 
veto on additional charged tracks and $\pi^0$'s. Thus there is no
information on the target particle and at some level events with $N^*$ 
production must be present. We know that the diffractive 
$\rho$ cross section at high energies is about 10 $\mu$b and the 
quasi-diffractive cross section with nucleon break-up
is about 5 $\mu$b. These, together with
the relative $f_2$ production rate, imply a cross section for the $f_2(1270)$
between $\sim 10$ nb and $\sim 30$ nb. That is, it is of the same order of
magnitude as our calculated $f_2$ cross section from odderon exchange.

Let us look first at the alternative explanation, namely Regge exchange. In 
a diffractive $\pi^+\pi^-$ photoproduction experiment at $\surd{s} \sim 6$ 
GeV the SLAC Hybrid Photon Collaboration \cite{SLAC} saw no evidence for
the $f_2(1270)$ in their $\pi^+\pi^-$ mass spectrum. The data in this
experiment correspond to genuine elastic diffraction with no quasi-elastic
events present in the final sample which was selected on the basis of a
three-constraint kinematic fit. If the E687 $f_2$ signal were due to
Regge exchange then, as we expect this cross section to vary approximately
as $s^{-1}$,  the $f_2$ signal at $\surd{s} \sim 6$ GeV would be $\sim 40$ 
to $\sim 120$ nb. This is comparable to the $\rho'$ cross section of 
$\sim 130$ nb and would have been visible in the experiment even at the 
lower limit. Of course, as explained earlier,  we would not expect any odderon 
contribution to be
seen in this experiment as nucleon break-up is specifically excluded.

In a similar experiment the CERN Omega Photon Collaboration \cite{CERN} at 
$\langle \surd{s} \rangle \sim 8.5$ GeV has also seen no evidence for the 
$f_2$ in their $\pi^+\pi^-$mass spectrum. The CERN experiment has lower 
statistics than the SLAC experiment, and the estimated Regge cross section 
is lower, lying between $\sim 20$ and $\sim 60$ nb. The upper end of this 
range can reasonably be ruled out but not the lower. The CERN experiment 
also differs from the SLAC experiment in that it cannot exclude some 
contamination from quasi-elastic events with nucleon break-up, so odderon 
exchange is allowed. Without a re-analysis of the data it is difficult to 
say categorically whether a cross section of 10 to 20 nb for the $f_2$
would be observable, although it does seem improbable.

Thus we can rather safely rule out Regge exchange as an explanation of the
$f_2$ signal in the E687 experiment, but can retain the possibility of
odderon exchange as an alternative to the ``missing-particles'' hypothesis. 
It may be purely coincidental that the experimental cross section for the 
$f_2$ is of the same order of magnitude as our calculation, but it is 
encouraging none the less. An additional and very relevant fact is that
the $t$-dependence of the odderon term is significantly less strong than
in diffractive $\rho$ photoproduction, for example. This means that the 
$f_2$ cross section
will appear to be relatively suppressed at small $t$, in conformity with
the E687 data. Of course none of this is sufficient to claim that the
odderon has been observed but it is sufficient to justify a new experimental
study to confirm or deny this hypothesis.

The $t$-dependence of the cross sections for $f_2$ and $\pi^0$ are very
similar, but the $Q^2$ dependence of the former is much less strong than
for the latter. Thus although the $f_2$ cross section is a factor of
about ten smaller than the $\pi^0$ cross section for production by real 
photons the difference is much less marked if a range of photon virtualities
is considered and $f_2$ production is competitive.
 
The experiments H1 and ZEUS at HERA are ideally suited to clarify the
situation since they have higher c.m. energy and can trigger on nucleon 
break-up. The challenge is then to observe a cross section of the order of
0.3 nb. for ${\rm e~p} \rightarrow {\rm e}f_2{\rm X}$.  

\section{Acknowledgements}

The authors would like to thank 
G. Kulzinger,
P. V. Landshoff,
P. Lebrun,
K. Meier,
H. J. Pirner,
K. D. Rothe and
M. Rueter
for useful discussions and correspondence.

{\it Supported in part by German 
Bundesministerium f\"ur Bildung und Forschung (BMBF),
Contract Nr. 05 7HD 91 P(0); 
by the EU Programme ``Training and Mobility
of Researchers'', Networks ``Hadronic Physics with High Energy 
Electromagnetic Probes'' (contract FMRX-CT96-0008);
by  ``Quantum 
Chromodynamics and the Deep Structure of Elementary Particles'' (contract
FMRX-CT98-0194); by PPARC; by DAAD and by the EU programme
``Human potential and mobility'', Contract Nr. HPMF-CT-1999-00179.}

\newpage

\begin{appendix}

\renewcommand{\theequation}{A.\arabic{equation}}
\setcounter{equation}{0}

\section{The decay of the $f_2$ into two photons}

In this appendix we calculate the partial width of the decay
$f_2 \rightarrow \gamma \gamma$
for the $f_2$ being in a
definite helicity state, 
$\Gamma(f_2(\lambda_f) \rightarrow \gamma \gamma)$.
We perform the calculation in the infinite
momentum frame, where the three momentum of the $f_2$ is very large,
$p_f^3 \rightarrow \infty$, and $\vec{p}_{f_T}=0$.

The starting point of the following quark model calculation is the
S-matrix element:
\begin{eqnarray}
 \label{klappt}
  && S_{fi} = \delta_{fi} + i (2 \pi)^4
  \delta^4(\sum_f p_f - \sum_i p_i) T_{fi}, \\
%  \nonumber\\
  &&T_{fi}=\langle \gamma(k_1,\lambda_1), \gamma(k_2,\lambda_2) | T
  | f_2(p_f,\lambda_f) \rangle := 
  T(k_1,k_2;\lambda_1,\lambda_2;\lambda_f)= 
  \nonumber\\
  && \int \frac{d^2 k_T}{(16 \pi)^3} \, \frac{1}{\sqrt{z \bar{z}}} \,
  \sum_{q,h,\bar{h}}
  \tilde{\Psi}^f_{\lambda_f,qh\bar{h}} 
%\nonumber\\
%  &&\hphantom{\int \frac{d^2 k_T}{(16 \pi)^3}} \times
  {\textstyle  \sqrt{\frac{1}{3}}}   
  \delta_{A\bar{A}} 
  \; \langle \gamma(k_1,\lambda_1), \gamma(k_2,\lambda_2) | T |
  q(p_q,h,A), \bar{q}(p_{\bar{q}},\bar{h},\bar{A}) \rangle.
\nonumber
% \label{klappt}
\end{eqnarray}
We calculate the partonic S-matrix element 
in (\ref{klappt}) 
%
%
%
%%%%%%%%%%%%%%%%%%%%%%%%%%%%%%%%% FIGURE
\begin{figure}[htb]
  \unitlength1.0cm
  \begin{center}
    \begin{picture}(15.,3.8)

      \put(1.0,1.0){
        \epsfysize=3.0cm
        \epsffile{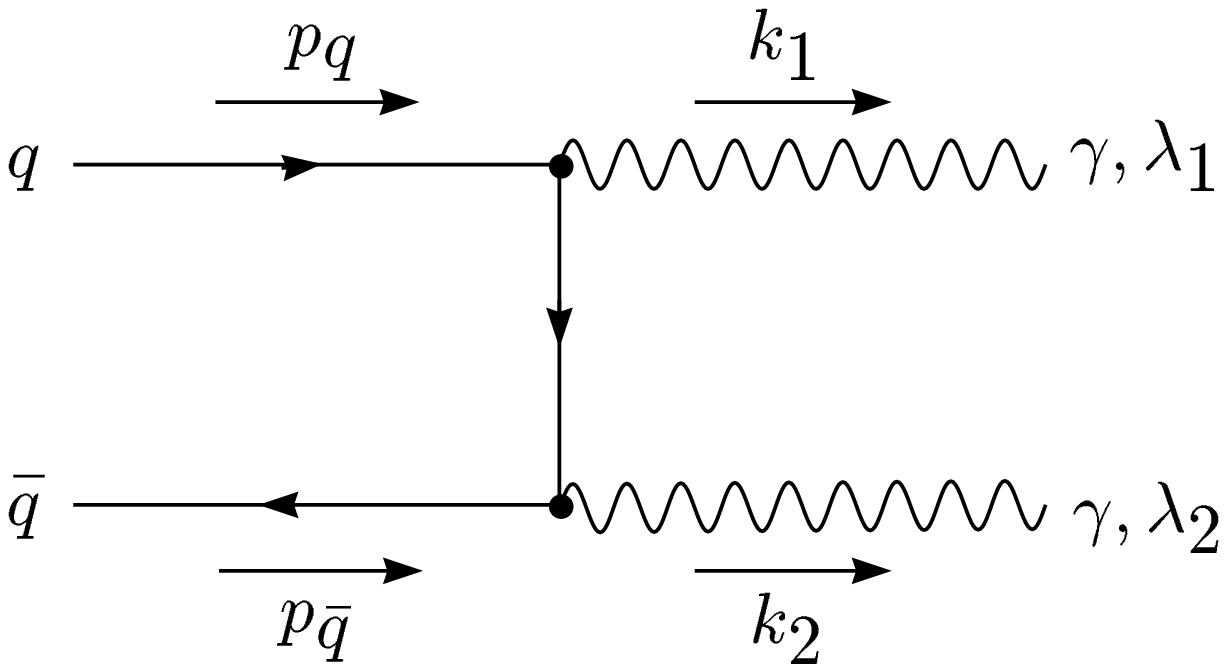}}

      \put(7.7,1.0){
        \epsfysize=3.0cm
        \epsffile{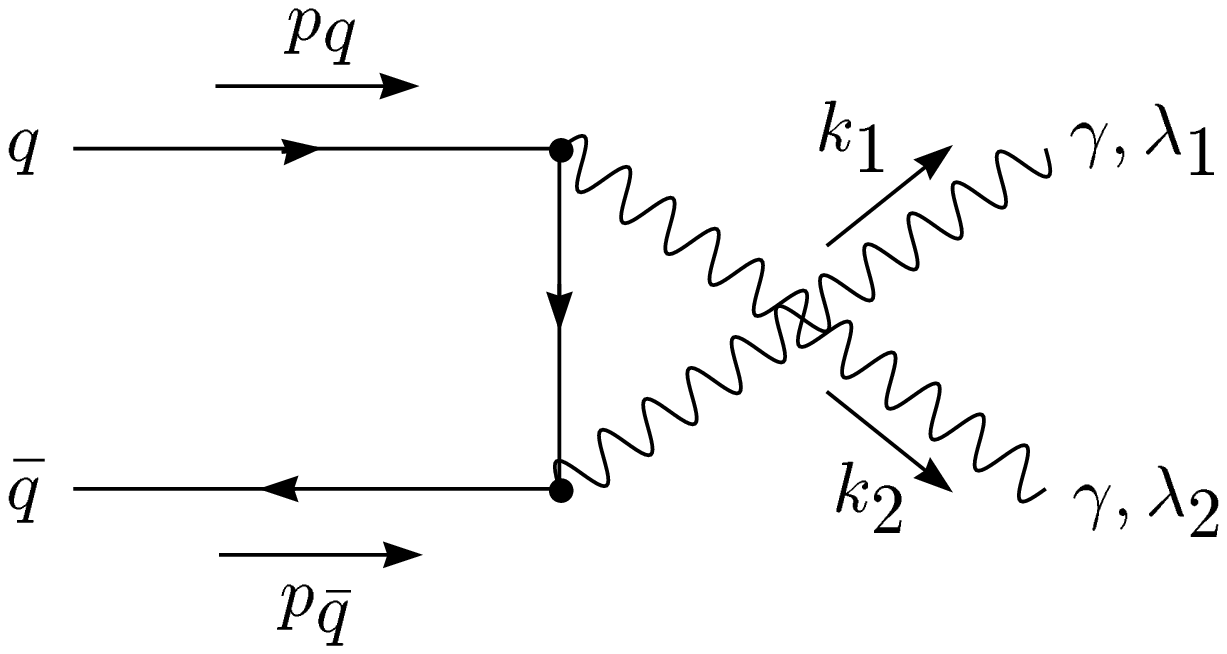}}

    \end{picture}
  \end{center}
  \vspace*{-2.0cm}
  \caption{The Feynman diagrams for the partonic S-matrix element 
in (\ref{klappt})}
\label{ddiagram}
\end{figure}
%%%%%%%%%%%%%%%%%%%%%%%%%%%%%%%%%%%%
%
%
%
in lowest
order perturbation theory. 
The Feynman diagrams are shown in
Figure \ref{ddiagram}. 
The corresponding analytic expression is:
\begin{eqnarray} 
  &&\langle \gamma(k_1,\lambda_1), \gamma(k_2,\lambda_2) | T |
  q(p_q,h,A), \bar{q}(p_{\bar{q}},\bar{h},\bar{A}) \rangle = 
\nonumber\\
  &&\hphantom{sss}\times
  \delta_{A\bar{A}} \, Q_q^2 e^2 \,
  \bar{v}(p_{\bar{q}},\bar{h}) \Big\{ \;
  {e\hskip-5.4pt /\hskip2pt}
  (\lambda_2) 
  \frac{ ( 
  {p \hskip-4.9pt /\hskip0pt}_q - 
  {k \hskip-5.9pt /\hskip0pt}_1 
  + m) }
  {(p_q-k_1)^2 -m^2 + i\epsilon}
  {e\hskip-5.4pt /\hskip2pt} 
  (\lambda_1) + 
\nonumber\\
  &&\hphantom{\; sss\times \delta_{A\bar{A}} \, Q_q^2 e^2 \,
  \bar{v}(p_{\bar{q}},\bar{h}) \Big\{ }
  {e \hskip-5.4pt /\hskip2pt}
  (\lambda_1) \frac{ ( 
  {p \hskip-4.9pt /\hskip0pt}_q -
  {k \hskip-5.9pt /\hskip0pt}_2  
   + m) }
  {(p_q-k_2)^2 -m^2 + i\epsilon} 
  {e\hskip-5.4pt /\hskip2pt}
  (\lambda_2) \; \; \Big\} u(p_q,h)
 \label{dpmatrix}
\end{eqnarray}
Here $Q_q$ is the quark charge. Combining (\ref{klappt}) 
and
(\ref{dpmatrix}) 
the decay rate of the $f_2$ with 
helicity $\lambda_f$ is given by
\begin{eqnarray} 
  &&\Gamma(f_2(\lambda_f) \rightarrow \gamma \gamma) = 
  \int \frac{1}{2 m_f} 
  (2 \pi)^4 \delta^4( p_f - k_1 - k_2)  
  \nonumber\\
  && \hphantom{d \Gamma(f_2(\lambda_f)}
  \times
  \frac{1}{2} 
  \frac{d^3k_1}{(2\pi)^3 2 k_1^0}
  \frac{d^3k_2}{(2\pi)^3 2 k_2^0}
  | T(k_1,k_2;\lambda_1,\lambda_2,\lambda_f) |^2
 \label{decay}
\end{eqnarray}
where we have made the approximation $p_q+p_{\bar{q}} \approx p_f$.
The factor $1/2$ reflects that the final state photons are  
indistingiushable. 

Now, since we consider the decay in the infinite momentum frame
of the $f_2$ it is useful to introduce light cone variables for the
photon momenta $k_1,k_2$ when performing the phase space integral in
(\ref{decay}),
\begin{eqnarray} 
  &&\frac{d^3k_1}{(2\pi)^3 2 k_1^0}
  \frac{d^3k_2}{(2\pi)^3 2 k_2^0}
  \delta^4( p_f - k_1 - k_2)=
  \nonumber\\
  &&\hphantom{sssss}
{\textstyle \frac{1}{2}} \prod_{i=1}^2 \Big[
  dk_{i_+} dk_{i_-} d^2k_{i_T} \,
  \frac{1}{k_{i_+}} \delta(k_{i_-} - \frac{k_{i_T}^2}{k_{i_+}})
  \Big]
  \nonumber\\
  &&\hphantom{sssss} 
   \delta (p_{f_+} - k_{1_+} - k_{2_+} )
   \delta (p_{f_-} - k_{1_-} - k_{2_-} )
   \delta^2 (\vec{k}_{1_T} +\vec{k}_{2_T} )
 \label{space}
\end{eqnarray}
From (\ref{space}) one can see, that we are left with an integral
over the transverse momentum of one photon, say
$\vec{k}_{1_T}:=\vec{q}_T$. Using (\ref{space}) we finally get for the 
partial decay width:
\begin{eqnarray} 
  &&\Gamma(f_2(\lambda_f) \rightarrow \gamma \gamma) = 
  \frac{1}{8 m_f} \frac{1}{(2 \pi)^2} \,
  \sum_{j=+,-} \int d^2q_T  
  \nonumber\\
  &&\hphantom{\Gamma(f_2(}\times
  \frac{1}{m_f^2}
  \frac{1}{\sqrt{1-4 q_T^2/m_f^2}} \, 
  \sum_{\lambda_1,\lambda_2} 
  | T_j(\vec{q}_T,\lambda_1,\lambda_2) |^2
 \label{decayres}
\end{eqnarray}
where we have to replace the plus, minus and transverse components of the
photon momenta in (\ref{decay}) according to (\ref{space}) by:
\begin{eqnarray} 
  &&\vec{k}_{1_T} = -\vec{k}_{2_T}=\vec{q}_T, \;
  k_{i_-} = \frac{q_T^2}{k_{i_+}},
  \nonumber\\
  &&k_{1_+} = 
  {\textstyle \frac{1}{2}}
  p_{f_+} 
  \big(1 \pm \sqrt{1-\frac{4 q_T^2}{m_f^2}}):=q_{\pm},
  \nonumber\\
  &&k_{2_+} = 
  {\textstyle \frac{1}{2}}
  p_{f_+} 
  \big(1 \mp \sqrt{1-\frac{4 q_T^2}{m_f^2}}):=q_{\mp},
 \label{replace}
\end{eqnarray}
Now we discuss (\ref{decayres}) in more detail: First there occurs
a sum over amplitudes $T_{\pm}$. For definite helicities of the photons
they are defined as
\begin{eqnarray}
  &&T_{+} : = T(k_{1_+} \rightarrow q_+,k_{2_+} \rightarrow q_-),
  \nonumber\\
  &&T_{-} : = T(k_{1_+} \rightarrow q_-,k_{2_+} \rightarrow q_+).
 \label{tplus}
\end{eqnarray}
where $T$ are the amplitudes defined in (\ref{klappt}).
When translating them into the rest frame
they correspond to the cases that either photon 1 ($+$) or photon 2
($-$) has positive momentum in three direction. This becomes clear by
looking at the expressions for $k_{i_+}$ in (\ref{space}). The ($+$)
case corresponds to the upper signs and the ($-$) case to the lower
ones. 

Second, again by considering the decay in the rest frame it is clear,
that no soft photon singularities can arise since the absolute value of
the photon momenta is $m_f/2$. In principle a very soft photon can be
produced by the boost into the infinite momentum frame when the photons
have only momentum in three direction. However, this special kinematical
situation corresponds to $|\vec{q}_T|:=q_T=0$ 
in (\ref{decayres}) and is supressed
due to a factor $q_T$ from the integration measure. 
Finally the sum over all photon polarisations does only
depend on $q_T$ and so the angle integration 
can be done trivially.

To calculate (\ref{decayres}) we generate 
for every photon
polarisation pair $(\lambda_1,\lambda_2)$
explicit expressions for the
matrix elements (\ref{klappt}) using mathematica and calculate
the integrals by numerical integration. After performing the sum over
the photon polarisations in (\ref{decayres}) we are left with a one
dimensional integral over $q_T$, whose value depends on the frequency
$\omega_{\lambda_f}$. This integral 
$\Gamma(f_2(\lambda_f) \rightarrow \gamma \gamma)$ of (\ref{decayres})
is plotted 
in Fig. \ref{omega} for the case $\lambda_f=2$ as a function of 
$\omega_{2}$.

\renewcommand{\theequation}{B.\arabic{equation}}
\setcounter{equation}{0}

\section{The Calculation of the $\gamma^{\ast}f_2$ overlap functions}

In this appendix we calculate the Fourier transforms
of the helicity wave functions (\ref{kresult}) and the
photon-tensor meson overlap functions.
We start by defining the configuration space helicity wave functions: 
\begin{eqnarray} 
  \Psi^f_{\lambda_f,qh\bar{h}}(\vec{r}_1,z) : =
  \int \frac{d^2k_T}{(2 \pi)^2} e^{i \vec{k}_T \vec{r}_1}
  \tilde{\Psi}^f_{\lambda_f,qh\bar{h}}(\vec{k}_T,z)
 \label{fourier}
\end{eqnarray}
The $\vec{k}_T$-dependence of the helicity structure of $\tilde{\Psi}$
can be expressed by spatial derivatives acting on the exponential in
(\ref{fourier}) and so, besides differentiating, it remains to calculate
the Fourier transforms $f_{\lambda_f}$ of the BSW functions (\ref{bsw})
with the result
\begin{eqnarray} 
  f_{\lambda_f}(r_1,z) = 
  N_{\lambda_f} \, (z \bar{z})^{3/2} \,
  e^{-\frac{1}{2} m_f^2 (z-1/2)^2 / \omega_{\lambda_f}^2} \times 
  \frac{1}{m_f} 
  e^{-\frac{1}{2} r_1^2 \, \omega_{\lambda_f}^2}.
 \label{bswkonf}
\end{eqnarray}
In this way we find for the helicity wave functions 
of a $q\bar{q}$-dipole (\ref{kresult}) in configuration
space ($c_u = c_d = 1/\sqrt{2}$): 
\begin{eqnarray} 
  && \Psi^f_{\pm2,qh\bar{h}} = (\pm2) \Big\{ i m(Q^2) e^{\pm i \theta_1} 
  \omega_2^2 r_1 \delta_{\pm \pm} \mp
  e^{\pm2 i \theta_1} \omega_2^4 r_1^2 
  (z \delta_{\pm \mp} - \bar{z} \delta_{\mp \pm})
  \Big\} c_q f_2,
  \nonumber\\
  && \Psi^f_{\pm 1,qh\bar{h}} = (-1) \Big\{ m_f m(Q^2) 
  (z-\bar{z}) \delta_{\pm \pm} \mp
\nonumber\\
  &&\hphantom{\Psi_{+1,qh\bar{h}} = (-1) \Big\}}
  i m_f e^{ \pm i \theta_1} \omega_1^2 r_1 
  \big( (3z - 4 z^2) \delta_{\pm \mp} +(3\bar{z} 
  - 4 \bar{z}^2) \delta_{\mp \pm} \big) \Big\}
  c_q f_1,
  \nonumber\\
  && \Psi^f_{\, 0 \, ,qh\bar{h}} = (
{\textstyle -\sqrt{\frac{2}{3}}}
) \Big\{ 
  i m(Q^2) \omega_0^2 r_1 (
  e^{-i\theta_1} \delta_{++} - e^{i\theta_1} \delta_{--}) +
\nonumber\\
  &&\hphantom{\Psi_{\, 0 \, ,qh\bar{h}} = 
{\textstyle -\sqrt{\frac{2}{3}}}}
  \big(2 z\bar{z} m_f^2 - \omega_0^4 r_1^2 + 2 \omega_0^2\big)
  \big( (z-\bar{z}) \delta_{+-} - (\bar{z}-z)\delta_{-+} \big)
  \Big\}
  c_q f_0,
 \label{konfwave}
\end{eqnarray}
Next we use (\ref{konfwave}) together with the 
photon wave functions as given in (4)-(6) of
\cite{doku} and calculate the $\gamma^{\ast}f_2$ 
overlap functions. With 
$\langle e \rangle = e/\sqrt{18}$ 
we find for transversly polarised photons:
\begin{eqnarray}  
  &&  \sum_{q,h,\bar{h}} 
  \Psi^{\ast f}_{+2,qh\bar{h}}
  \Psi^{\gamma}_{1,qh\bar{h}}= 
  i e^{-i \theta_1} (-2) \sqrt{6} \langle e \rangle 
  \Big\{ 
  m(Q^2)^2  \omega_2^2 r_1 \frac{K_0(\epsilon r_1)}{2 \pi} +
\nonumber\\
  &&\hphantom{\sum_{q,h,\bar{h}} 
  \Psi^{\ast}_{+2,qh\bar{h}}
  \Psi^{\gamma}_{1,qh\bar{h}}=} 
  \omega_2^4 r_1^2 
  (z^2 + \bar{z}^2)
  \epsilon \frac{K_1(\epsilon r_1)}{2 \pi} \Big\}
  f_2(r_1,z)
  \nonumber\\
  &&  \sum_{q,h,\bar{h}} 
  \Psi^{\ast f}_{-2,qh\bar{h}}
  \Psi^{\gamma}_{1,qh\bar{h}}= 
  i e^{3 i \theta_1} (2)  \sqrt{6} \langle e \rangle \Big\{ 
  \omega_2^4 r_1^2 
  (2 z\bar{z})
  \epsilon \frac{K_1(\epsilon r_1)}{2 \pi} \Big\}
  f_2(r_1,z)
\nonumber\\
  &&  \sum_{q,h,\bar{h}} 
  \Psi^{\ast f}_{+1,qh\bar{h}}
  \Psi^{\gamma}_{1,qh\bar{h}}= 
  (-1) \sqrt{6} \langle e \rangle \Big\{
  m_f m(Q^2)^2 (z-\bar{z}) \frac{K_0(\epsilon r_1)}{2 \pi} +
\nonumber\\
  &&\hphantom{\sum_{q,h,\bar{h}} 
  \Psi^{\ast}_{+1,qh\bar{h}}
  \Psi^{\gamma}_{1,qh\bar{h}}= }
  m_f \omega_1^2 r_1 (z-\bar{z})(1-4 z\bar{z}) 
  \epsilon \frac{K_1(\epsilon r_1)}{2 \pi} \Big\}
  f_1(r_1,z)
\nonumber\\
  &&  \sum_{q,h,\bar{h}} 
  \Psi^{\ast f}_{-1,qh\bar{h}}
  \Psi^{\gamma}_{1,qh\bar{h}}= 
  e^{2 i \theta_1}  \sqrt{6} \langle e \rangle \Big\{
  m_f \omega_1^2 r_1 (\bar{z}-z)(4 z\bar{z}) 
  \epsilon \frac{K_1(\epsilon r_1)}{2 \pi} \Big\}
  f_1(r_1,z)
\nonumber\\
  &&  \sum_{q,h,\bar{h}} 
  \Psi^{\ast f}_{\, 0 \,,qh\bar{h}}
  \Psi^{\gamma}_{1,qh\bar{h}}= 
  i e^{i \theta_1} 2 
  \langle e \rangle \Big\{
  m(Q^2)^2 \omega_0^2 r_1 
  \frac{K_0(\epsilon r_1)}{2 \pi} -
\nonumber\\
  &&\hphantom{\sum_{q,h,\bar{h}} 
  \Psi^{\ast}_{\, 0 \,,qh\bar{h}}
  \Psi^{\gamma}_{1,qh\bar{h}}= }
  (2 z\bar{z} m_f^2 - \omega_0^4 r_1^2 + 2 \omega_0^2)
  (z - \bar{z})^2  
  \epsilon \frac{K_1(\epsilon r_1)}{2 \pi} \Big\}
  f_0(r_1,z)
 \label{transov1}
\end{eqnarray}
The $\gamma^{\ast} f_2$ overlap functions for longitudinally polarised
photons are:
\begin{eqnarray}
  &&  \sum_{q,h,\bar{h}} 
  \Psi^{\ast f}_{+2,qh\bar{h}}
  \Psi^{\gamma}_{0,qh\bar{h}}=
  Q e^{-2 i \theta_1} 2 \sqrt{3} \langle e \rangle \Big\{
  \omega_2^4 r_1^2 (z - \bar{z}) (2 z \bar{z}) 
  \frac{K_0(\epsilon r_1)}{2 \pi}  \Big\}
  f_2(r_1,z),
  \nonumber\\
  &&  \sum_{q,h,\bar{h}} 
  \Psi^{\ast f}_{+1,qh\bar{h}}
  \Psi^{\gamma}_{0,qh\bar{h}}= 
  Q i e^{-i \theta_1} \sqrt{3} \langle e \rangle \Big\{
  m_f \omega_1^2 r_1 (2 z \bar{z})
\nonumber\\
  &&\hphantom{\sum_{q,h,\bar{h}} 
  \Psi^{\ast}_{-1,qh\bar{h}}
  \Psi^{\gamma}_{-1,qh\bar{h}}=} \times
  \big( (3z-4z^2) +(3\bar{z}-4\bar{z}^2) \big)
  \frac{K_0(\epsilon r_1)}{2 \pi}  \Big\}
  f_1(r_1,z),
\nonumber\\
  &&\sum_{q,h,\bar{h}} 
  \Psi^{\ast f}_{0,qh\bar{h}}
  \Psi^{\gamma}_{0,qh\bar{h}}= 
  Q \sqrt{2} \langle e \rangle \Big\{
  (2 z\bar{z} m_f^2 - \omega_0^2 r_1^2 + 2 \omega_0^2)
\nonumber\\
  &&\hphantom{\sum_{q,h,\bar{h}} 
  \Psi^{\ast}_{0,qh\bar{h}}
  \Psi^{\gamma}_{0,qh\bar{h}}= 
  Q \sqrt{2}}\times
  (2 z\bar{z})2(z-\bar{z}) 
  \frac{K_0(\epsilon r_1)}{2 \pi}  \Big\}
  f_0(r_1,z),
 \label{longov}
\end{eqnarray}
Finally the remaining overlap functions are simply related to the
overlap functions (\ref{transov1}), (\ref{longov}) by:
\begin{eqnarray}   
  \sum_{q,h,\bar{h}} 
  \Psi^{\ast f}_{+\lambda_f,qh\bar{h}}
  \Psi^{\gamma}_{+\lambda_{\gamma},qh\bar{h}}=
  \Big(\sum_{q,h,\bar{h}} 
  \Psi^{\ast f}_{-\lambda_f,qh\bar{h}}
  \Psi^{\gamma}_{-\lambda_{\gamma},qh\bar{h}} \Big)^*,
 \label{transov2}
\end{eqnarray}

\end{appendix}


\newpage

\begin{thebibliography}{99}


\bibitem{nico1}
  L Lukaszuk and B Nicolescu: Nuov. Cim. Lett. {\bf 8}, 405 (1973);\\
  D Joynson, E Leader, C Lopez and B Nicolescu: Nuov. Cim.
  {\bf A30}, 345  (1975)

\bibitem{etal}
  E R Berger, A Donnachie, H G Dosch, W Kilian, 
  O Nachtmann and M Rueter: Eur. Phys. J. {\bf C9}, 491 (1999)

\bibitem{doruod}
  M Rueter and H G Dosch: Phys. Lett. {\bf B380},  177 (1996)

\bibitem{donaru}
  M Rueter, H G Dosch and O Nachtmann:
  Phys. Rev. {\bf D59}, 014018 (1999)

\bibitem{nico2}
  B Nicolescu: report hep-ph/9911334

\bibitem{na91}
  O Nachtmann: Ann. Phys. {\bf 209}, 436 (1991)

\bibitem{dfk}
  H G Dosch, E Ferreira and A Kr\"amer: 
  Phys. Rev. {\bf D50}, 1992 (1994)

\bibitem{all}
  O Nachtmann: in {\it Perturbative and 
  Nonperturbative Aspects of Quantum Field Theory},
  edited by H Latal and W Schweiger
  (Springer Verlag,
  Berlin, Heidelberg 1997)

\bibitem{lightcone}
  J D Bjorken, J B Kogut and D E Soper:
  Phys. Rev. {\bf D3}, 1382 (1971);\\
  G P Lepage and S J Brodsky:
  Phys. Rev. {\bf D22}, 2157 (1980);\\
  G P Lepage: in {\it Proceedings of the Banff
  Summer Institut, Particles and Fields 2}, Banff (Canada 1982) 

\bibitem{dogupi}
  H G Dosch, T Gousset and H J Pirner: Phys. Rev.
  {\bf D57}, 1666 (1998)


\bibitem{guku}
  H G Dosch, T Gousset, G Kulzinger and H J Pirner: Phys. Rev.
  {\bf D55}, 2602 (1997)

\bibitem{bsw}
  M Wirbel, B Stech and M. Bauer: Z. Phys. {\bf C29}, 637 (1985)

\bibitem{particle} R M Barnett et al (Particle Data Group):
  Eur. Phys. J. {\bf C3}, 1 (1998)

\bibitem{epa} V M Budnev, I F Ginzburg, G V Meledin and
  V G Serbo: Phys. Rep. {\bf 15}, 181  (1975)
 
\bibitem{E687} P Lebrun: {\it Proceedings of HADRON'97} p.504, edited by 
  S-U Chung and H J Willetski (AIP Conference Proceedings 432, 1997);\\ 
  E687: private communication

\bibitem{SLAC} K. Abe et al: Phys.Rev.Lett. {\bf 53} (1984) 751

\bibitem{CERN} D Aston et al: Phys.Lett. {\bf 92B} (1980) 215

\bibitem{doku}
  G Kulzinger, H G Dosch and H J Pirner: 
  Eur. Phys. J. {\bf C7}, 73 (1999)

\end{thebibliography}
\end{document}